\documentclass[pdflatex,sn-mathphys-num]{sn-jnl}
\usepackage{graphicx}%
\usepackage{multirow}%
\usepackage{amsmath,amssymb,amsfonts}%
\usepackage{amsthm}%
\usepackage{mathrsfs}%
\usepackage[title]{appendix}%
\usepackage{xcolor}%
\usepackage{textcomp}%
\usepackage{manyfoot}%
\usepackage{booktabs}%
\usepackage{algorithm}%
\usepackage{algorithmicx}%
\usepackage{algpseudocode}%
\usepackage{lineno,ulem}
\usepackage{listings,overpic}%
\theoremstyle{thmstyleone}%

\theoremstyle{thmstyletwo}%
\theoremstyle{thmstylethree}%
\begin{document}

\title[Article Title]{Harnessing Measurement Error as a Thermodynamic Resource in Two-Reservoir Information Engines}

\author[1]{\fnm{Yang} \sur{Xiao}}
\author[2]{\fnm{Qian} \sur{Zeng}}

\author*[3]{\fnm{Jin} \sur{Wang}}\email{jin.wang.1@stonybrook.edu}

\affil[1]{\orgdiv{College of Physics}, \orgname{Jilin University,} \orgaddress{\city{Changchun,} \postcode{130022}, \country{China}}}

\affil[2]{\orgdiv{State Key Laboratory of Electroanalytical Chemistry}, \orgname{Changchun Institute of Applied Chemistry}, \orgaddress{\city{Changchun}, \postcode{130022},  \country{China}}}

\affil*[3]{\orgdiv{Department of Chemistry and Department of Physics and Astronomy}, \orgname{State University of New York at Stony Brook}, \orgaddress{\street{Stony Brook}, \city{New York}, \postcode{11794},  \country{USA}}}

\abstract{In general, measurement error constrains the performance of information engines. However, this may not hold universally. 
In this work, we propose a class of two-reservoirs information engines (TRIEs) with a two-level system  and  demonstrate that the monotonic relationships between the work output, the efficiency,  as well as the work fluctuations  of the TRIEs and measurement error are determined  by  the demon's control parameters. 
Utilizing this, we establish the conditions that allow the TRIEs to generate useful work under any level of measurement error, and identify the regimes in which these performance parameters can be optimized simultaneously by the measurement error.
{Furthermore, we demonstrate that when the demon’s energy cost is supplied solely by the cold reservoir, the efficiency that accounts for this cost can surpass the standard Carnot bound without violating the second law}. 
Finally,  we design a dual  Otto cycles TRIE and provide an experimental scheme for its implementation based on a trapped $^{40}\mathrm{Ca}^+$ ion. Significantly, our results  can be applied to the  multi-level system information engines.}

\keywords{Information engines, Measurement error, control parameters, Trapped ions}

\maketitle

\section{Introduction}\label{sec1}

Information engines, devices that extract work from heat reservoirs by utilizing information, represent a foundational pillar of information thermodynamics. These engines require a monitoring mechanism, often referred to as a "demon", to oversee the system's state and control its evolution across classical models \cite{Szilard1929,Sei12,Sag14,Sagawa20,Yan24,TK23}, quantum systems \cite{CE18,Lutz23,Sagawa08,ueda18}, and active matter \cite{PM22,LC24}, especially under noisy limits \cite{TK22,AT22,JD24}.
 By analyzing the exchange of entropy between the demon and the system,   these engines save the second law of thermodynamics  \cite{Szilard1929,Sagawa08,Zeng21,Landauer61,KM09}. 
 Beyond being merely theoretical constructs, information engines are also observed in diverse natural and experimental systems, 
including cellular computation \cite{GT16, EM21, Lan16, PM12, Wen18}, biochemical signaling \cite{Ito15, JS14, Tu08, HD24,tu12, IL10}, 
 molecular machines \cite{Sc24,VS07,SF23,AP2008,Sag10,GP18}, and micro-nano systems \cite{SS20, JV15,JV14,KC17,MD16,NC17,Naghiloo18}.
Thus, information engines are foundational to the field of information thermodynamics and possess significant practical applications.

A key challenge limiting their real-world applications is the unavoidable measurement error caused by thermal noise and inaccuracies in measurement devices, and so on. This error leads to discrepancies between the demon's memory and the true state of the system, resulting in incorrect control operations and reduced work extraction \cite{TK22, Sag14, Sei12, Sagawa20, JH14, JH2214, PS14, KV25, AT22, DAS24, LC24, JD24}. Therefore, measurement error is often considered a harmful factor in information engines.

{To mitigate the detrimental effects of the measurement error, existing optimization strategies can be broadly categorized into two approaches. 
The first focuses on optimizing the measurement process itself, including enhancing measurement precision \cite{Sagawa20,Sag14}, performing continuous measurement \cite{NE25,JP23,MR19}, and employing Bayesian filtering \cite{TK22, AT22, LC24, JD24}. While these methods improve information acquisition, they  not only cause significant costs for erasing  memory due to Landauer's principle \cite{Landauer61}, but also will reach a performance bottleneck  when the measurement becomes uninformative,  resulting in work output disappearing. The second approach focuses on optimizing the feedback mechanism, including optimizing feedback protocols \cite{Sei12,PS14,Seifert11,KT25,JM11,Ueda12},  delayed-feedback control \cite{KV25}.  While these approaches can achieve impressive performance, they are  vulnerable to high noise levels.}

{Based on the limitations of current methods}, we propose a paradigm shift.
Rather than viewing measurement error strictly as a detriment to be minimized, we investigate whether it is feasible to extract positive work under any degree of measurement error, or even to harness this error as a resource for  generating useful work. The key to achieving this objective lies in establishing a quantitative relationship between the output work of the information engines and the measurement error.

Enhancing the practical utility of information engines requires addressing not only work extraction but also two other critical factors: thermodynamic efficiency and work fluctuations,  reflecting the stability of the engines. An ideal information engine should exhibit significant work output, high efficiency (low heat dissipation), and high stability, as  emphasized in traditional heat engines without demon \cite{hol18, sei18,lutz21}. Achieving this demands that these performance indicators demonstrate consistent trends of change within a certain parameter regime.

To address these challenges, we introduce a class of two-reservoir information engines (TRIEs) with a two-level working system. We demonstrate that the quantitative relationships between the output work,  efficiency, as well as  stability of the TRIEs and measurement error are determined by the control parameters of the demon. Utilizing this result, we show that the TRIEs can produce useful work under any measurement error,  {and} provide mechanisms for optimizing these performance parameters simultaneously using measurement error. 
{Furthermore, the  global efficiency of the TRIEs, taking  accounting  of the demon's cost,  can go beyond the standard Carnot efficiency.}
Finally, to illustrate our theoretical results, we propose a dual Otto cycles TRIE based on the standard Otto cycle \cite{Quan07, TD2004}, which comprises two adiabatic and two isochoric strokes, and present an experimental scheme for its implementation based on a trapped $^{40}\mathrm{Ca}^+$ ion. 
Importantly, our findings are not confined to two-level systems but can be extended to multi-level information engines.

\section{Results}\label{sec2}
\subsection{Model}\label{subsec2}

We establish a class of two-reservoir information engines (TRIEs) consisting of a two-level system and a demon. In this system, the ground state energy is assumed to be zero, while the excited state energy is denoted as $E_c$.   Initially, the system is in thermal equilibrium with a Gibbs heat reservoir at an inverse temperature $\beta_c$. The probabilities for the system occupying the ground state and the excited state are given by  $p_{g}=1/[1+\exp(-\beta_cE_c)]$, and $p_e=\exp(-\beta_cE_c)/[1+\exp(-\beta_cE_c)]$, respectively. 

\begin{figure}[h]
\centering
\begin{overpic}[width=8cm]{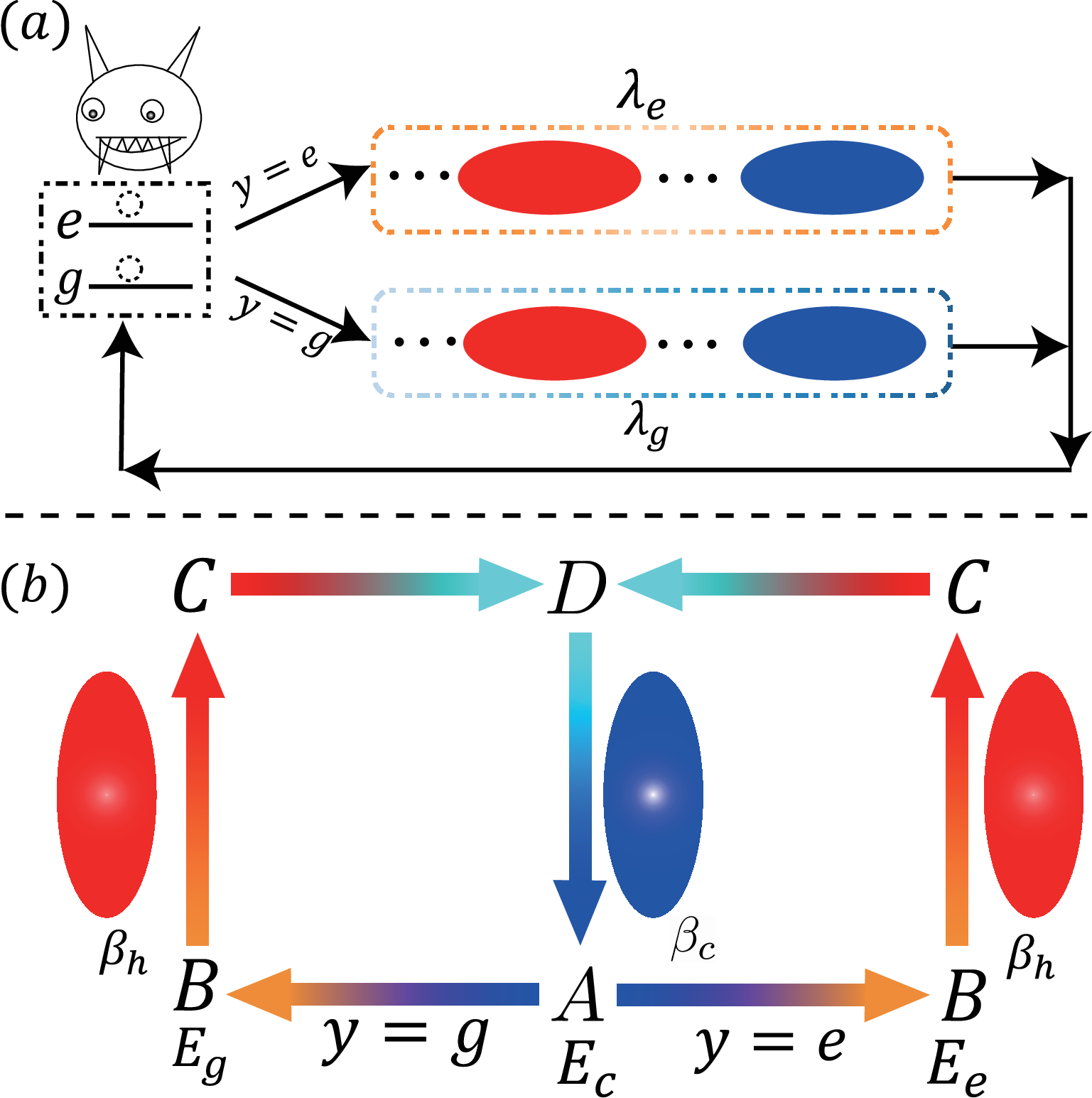}
\put(47,68){\scalebox{1.1}{$\beta_h$}}
\put(74,68){\scalebox{1.1}{$\beta_c$}}
\put(47,83){\scalebox{1.1}{$\beta_h$}}
\put(74,83){\scalebox{1.1}{$\beta_c$}}
\end{overpic}
\caption{(a) Schematic diagram of the TRIEs. (b) Operation flowchart of the dual Otto cycles TRIE. At point $A$,  the demon performs a measurement on the system.  Depending on the measurement outcome $y$,  the demon controls the system to evolve adiabatically from the initial energy gap $E_c$ to $E_y$, then couples the system to the hot reservoir at inverse temperature $\beta_h$ with fixed gap $E_y$.
    After thermalization, the system  is adiabatically returned to initial gap $E_c$, and
    coupled to a cold reservoir at inverse temperature $\beta_c$ with fixed frequency $E_c$ until it reaches its initial thermal distribution.}\label{Model}
\end{figure}

 At this moment, the demon makes a measurement of the system's state $x_i (x_i=e,g)$  with measurement outcome $y$.
 Based on this measurement outcome, the demon performs the corresponding control: If  $y=e$, the demon applies control 
 $\lambda_e$ on the system;  if $y=g$,  the demon  implements the control $\lambda_g$ on the system, as illustrated in Fig. \ref{Model}(a). Here, $\lambda_{y}$  represents the external parameter space (such as the energy gap of the system and the temperatures of the heat reservoirs) that affects the dynamic trajectory of the system.

 We impose three conditions on the demon's measurement and control operations: (1) Accuracy invariance: The measurement accuracy remains unchanged when measuring each energy level. Correspondingly, the measurement accuracy is defined as $p(y= x_i|x_i)=1-\epsilon$
 with $\epsilon$ representing the measurement error $p(y\neq x_i|x_i)=\epsilon$.  After the measurement, the probabilities for the demon's memory states 
  $y=e$ and $y=g$ can be expressed as:  $p(y=e)=p_e(1-\epsilon)+p_g\epsilon$ and $p(y=g)=p_e\epsilon+p_g(1-\epsilon)$, respectively.
The information $I=\sum_{x_i,y}p(y,x_i)[\ln(y|x_i)-\ln p(y)]$ obtained by the demon from the system can then be expressed as (see Methods \ref{M1} ‘The relations between the {local} efficiency and the entropy production’ )
\begin{eqnarray}\label{IAa}
I=S_Y-S_\epsilon.
\end{eqnarray}
where $S_Y=-p(y=e)\ln p(y=e)-p(y=g)\ln p(y=g)$  denotes the entropy of the demon's memory after the measurement, and $S_\epsilon=-(1-\epsilon)\ln(1-\epsilon)-\epsilon\ln\epsilon$ represents the entropy associated with the measurement error.
  (2) Each control operation can return the system to its initial distribution. This condition ensures that the system can operate cyclically and guarantees decoupling between the system and the demon at the end of one cycle.
 (3) Hot Reservoir Coupling:  During the control process, the demon couples the system with a hot reservoir at an inverse temperature $\beta_h$ ($0<\beta_h<\beta_c$). This arrangement is necessary {for introducing thermal driving force}  because  when $\beta_h=\beta_c$ there is no heat dissipation in the TRIEs due to energy conservation.

\subsection{Performance}

During a cycle, the  forward-time trajectory of the system is defined as $X_F$, which contains the initial state $x_i$.
Based on the initial state of the system and the measurement outcomes, the measurement operation can be classified into two categories: the correct measurement  $\mathcal{O}_{cor}$ (where $y=x_i$) and the erroneous measurement  $\mathcal{O}_{err}$
(where $y\neq x_i$).  The probabilities of measurement operations $\mathcal{O}_{cor}$ and $\mathcal{O}_{err}$ are given by $p(\mathcal{O}_{cor})=\sum_{x_i,y}p(y=x_i|x_i)p(x_i)=1-\epsilon$ and $p(\mathcal{O}_{err})=\sum_{x_i,y}p(y\neq x_i|x_i)p(x_i)=\epsilon$, respectively. Defining the stochastic work produced by the system along the trajectory $X_F$ as $w$, the average of the $m$-th power of the work conditioned on the correct and erroneous measurements can be expressed as: $\langle w_{cor}^m\rangle=\sum_{X_F,y}p(X_F|x_i,y)p(y|x_i)p(x_i)w^m(X_F,y)\delta_{x_i,y}/p(\mathcal{O}_{cor})=\sum_{X_F}p(X_F|x_i,y=x_i)p(x_i)w^m(X_F)$, and $\langle w_{err}^m\rangle=\sum_{X_F,y}p(X_F|x_i,y)p(y|x_i)p(x_i)w^m(X_F,y)(1-\delta_{x_i,y})/p(\mathcal{O}_{err})=\sum_{X_F,y}p(X_F|x_i,y)p(x_i)w^m(X_F,y)(1-\delta_{x_i,y})$,  respectively. Obviously, both $\langle w_{cor}^m\rangle$ and $\langle w_{err}^m\rangle$ are independent on measurement error and are dependent on the trajectory $X_F$ and the control $\lambda_y$.
Consequently, the average of the 
$m$-th power of the work $\langle w^m\rangle =p(\mathcal{O}_{cor})\langle w_{cor}^m\rangle+p(\mathcal{O}_{err})\langle w_{err}^m\rangle$ can be obtained
\begin{eqnarray}\label{wm}
   \langle w^m\rangle =(1-\epsilon)\langle w_{cor}^m\rangle+\epsilon \langle w_{err}^m\rangle.
\end{eqnarray}
Then the average work $\langle w\rangle$ can be   expressed as 
 \begin{align} \label{w}
\langle w\rangle&=\langle w_{cor}\rangle+\epsilon \langle w_d\rangle,
\end{align} 
where  $\langle w_d\rangle=\langle w_{err}\rangle-\langle w_{cor}\rangle$ quantifies the difference in the work generated by the system under conditions of erroneous measurement compared to correct measurement.

In the process of deriving Eq. (\ref{w}), we did not apply condition (3)  to the demon. Thus, Eq. (\ref{w}) also applies to single heat reservoir information engines (SRIEs). In Eq. (\ref{w}), the output work is a monotonic function of the measurement error, as $\langle w_d\rangle$ is independent on measurement error.
When $\langle w_d\rangle<0$ (where the system produces more work under correct measurement than under incorrect measurement {and the demon is a good demon}), the work $w$ decreases as increasing measurement error $\epsilon$. Conversely, when $\langle w_d\rangle>0$ (where the system produces less work under correct measurement than under incorrect measurement), the work $w$ increases as measurement error $\epsilon$ increases. Under this mechanism, the  measurement error is transferred to  a utilizable resource for work extraction from a detrimental factor.
Here, we imagine 'correct measurement' and 'incorrect measurement' as two different thermodynamic trajectories accessible to the demon. The control parameter $\lambda_y$ determines which trajectory produces higher work. The error $\epsilon$ is a knob that allocates the probability of system operation between these two trajectories. Therefore, if the wrong trajectory is more efficient, increasing the error is increasing the probability of the system using the efficient trajectory. {Although this design is usually a bad demon, it is actually very important for real-world information engines. When measurement errors are large and unavoidable, such as in noisy and heat fluctuation environments, the  traditional good demons that depend on highly accurate measurements will fail.
In contrast, by intentionally coupling the erroneous trajectory to a higher thermodynamic gradient, which  compensates for the lack of accurate information, the bad demon can enable the engine to produce positive work even in highly noisy environments.}

Furthermore, the zero point of Eq. (\ref{w}) is located at $\epsilon=\epsilon_0=-\langle w_{cor}\rangle/\langle w_d\rangle$. Based on this, we can determine the conditions under which the TRIEs can produce positive work regardless of measurement error: when $\langle w_d\rangle<0$, $\epsilon_0>1$ is required where $0<\langle w_{err}\rangle<\langle w_{cor}\rangle$; when $\langle w_d\rangle>0$,  $\epsilon_0<0$ is needed where $0<\langle w_{cor}\rangle<\langle w_{err}\rangle$. 
{In a word, the condition for the TRIEs to produce work under any measurement error is that both the incorrect  measurement  trajectories and the correct measurement trajectories produce positive work.}
However, this property is impossible for the  SRIEs with a two-level system that share the same measurement error, cold reservoir, and initial mutual information as the TRIEs for the following reasons: Assuming the work produced by the SRIEs is denoted as $W$, according to the Sagawa-Ueda theorem \cite{Sagawa08,sagawa12}, the relationship $W\leq I/\beta_c$ must hold, where $I/\beta_c$ represent the maximum extractable work from  the SRIEs. With reference to Eq. (\ref{IAa}), when $\epsilon=0.5$, the demon cannot extract information from the system  $I=0$,  resulting in  $W\leq 0$. Thus, for SRIEs, if $\langle w_d\rangle<0$, as the case in \cite{Sag14, KV25}, in the regime $\epsilon>0.5$, it follows that  $W<0$. Conversely, if $\langle w_d\rangle>0$, in the regime $\epsilon<0.5$, we again have $W<0$. Therefore, the SRIEs are constrained to produce positive work only in the regimes $0\leq\epsilon<0.5$ or  $0.5<\epsilon\leq 1$.

{Therefore, the general principles that enable an information engine to operate under any level of measurement error are as follows:(i) Accuracy invariance: The probability of correctly measuring each energy level is uniform (i.e., $p(y=x_i|x_i) = 1-\epsilon$), ensuring the linear relationship derived in Eq. (\ref{w}).
 (ii)Hot reservoir coupling: The engine is coupled to two heat reservoirs at different inverse temperatures $\beta_h$ and $\beta_c$, which provide a thermal gradient. This additional thermodynamic degree of freedom allows work extraction when the mutual information gain is small, i.e. when $I=0$, the TRIEs can be driven by thermal gradient force. 
 (iii)Positive output work: The most critical principle is that the demon's control parameters are configured such that  the correct and erroneous measurement branches yield positive work.}

The stochastic heat absorbed by the system from the hot reservoir along the trajectory $X_F$ is defined as  $q_{h}$. Using the method for determining the average work, the average of this heat can be expressed as
$\langle q_h\rangle=\langle q_{cor}\rangle+\epsilon \langle q_d\rangle$, where $\langle q_{cor}\rangle$  represents the average heat exchange between the system and the hot reservoir conditioned on the correct measurement. The term $\langle q_d\rangle=\langle q_{err}\rangle-\langle Q_{cor}\rangle$  quantifies the difference in heat absorption between incorrect and correct measurements, with $\langle q_{err}\rangle$ denoting the heat absorbed from the hot reservoir conditioned on the incorrect measurement.
The thermodynamic  efficiency, defined as the ratio of work output to heat injected into the system, can then be expressed as \cite{Sagawa08,TT25,Jar13}
\begin{eqnarray}\label{eff}
 \eta=\frac{\langle w_{cor}\rangle+\epsilon \langle w_d\rangle}{\langle q_{cor}\rangle+\epsilon \langle q_d\rangle}.
\end{eqnarray}
{Since $\eta$ only contains the heat exchange between the system and the thermal reservoir, without taking into account the cost of the demon,  it is a local efficiency.}
This {local} efficiency is also a monotonic function
of the measurement error due to the sign of its  derivative with respect to the measurement error
$\eta^{'}=d\eta/d\epsilon=(\langle w_d\rangle\langle q_{cor}\rangle-\langle w_{cor}\rangle \langle q_d\rangle)/\langle q_h\rangle^2$ being independent of the measurement error. 
When $\eta^{'}<0$, the measurement error degrades the {blue}{local} efficiency. 
Within this framework, we can determine the parameter conditions for the  TRIEs to achieve perfect {local} efficiency, a feat unattainable by traditional heat engines without demons due to inherent heat dissipation. Remarkably, by solving $\eta=1$,  the exact parameter regime for
100\% {local} efficiency can be determined $\epsilon=\epsilon_p=(\langle w_{cor}\rangle-\langle q_{cor}\rangle)/(\langle q_{d}\rangle-\langle w_{d}\rangle)$. In this measurement error regime, when $\epsilon<\epsilon_p<\epsilon_0$, the {local} efficiency in Eq. (\ref{eff}) will exceed one ($\eta>1$)
due to the positive heat absorption from the cold reservoir $\langle q_c\rangle=\langle w\rangle-\langle q_h\rangle>0$. At this stage,  the TRIEs's {local} efficiency, defined as the ratio of work output
to heat injected into the system,   should be expressed as $\eta =\langle w\rangle/(\langle q_h\rangle+\langle q_c\rangle)=1$.  Thus, the perfect {local} efficiency occurs when $\epsilon\leq\epsilon_p<\epsilon_0$. 
Conversely,  by strategically selecting control parameters to achieve  $\eta^{'}>0$, the measurement error  enhances {local} efficiency. Mirroring the analysis in the suppression regime, strategic parameter selection that satisfies {$\epsilon_0<\epsilon_p\leq\epsilon$} allows the {local} efficiency of the TRIEs to  reach  100\%. This indicates that measurement errors can be utilized to increase the {local} efficiency of the  TRIEs.

The regimes of  achieving perfect {local} efficiency  are allowed in information thermodynamics.
Upon completion of the cycle,  the system returns to its initial  distribution, resulting in zero entropy production and a decoupling from the demon. This decoupling indicates that the changes in mutual information between the system and the demon during the control strokes equal the initial mutual information $I$.
Thus, the total entropy production for the system, the demon, and the reservoirs can be expressed as  $\Sigma_{tot}=\delta S_r +I\geq0$ \cite{sagawa12} with $\delta S_r=\delta S_h+\delta S_c$ denoting the entropy change of the heat reservoirs. Here, $\delta S_h = -\beta_h \langle q_h\rangle$ and $\delta S_c = -\beta_c \langle q_c\rangle$ \cite{Crook99,Jar00} represent the entropy changes of the hot and cold reservoirs, respectively. 
Based on these two equations, 
the {local} efficiency of the TRIEs  can be expressed as $\eta=(\delta S_h/ \beta_h+\delta S_c/\beta_c)/(\delta S_h/\beta_h)=1+(\delta S_c/\beta_c-\delta S_h/\beta_c+\delta S_h/\beta_c)/(\delta S_h/\beta_h)$, which can be further simplified as
\begin{eqnarray}\label{eta1}
    \eta=\eta_C+\frac{I-\Sigma_{tot}}{\beta_c \langle q_h\rangle}.
\end{eqnarray}
Since $\Sigma_{tot} \geq 0$, it follows that $\delta S_r\geq -I$. Moreover, given that $I \geq 0$,  the entropy change   of the reservoirs can satisfy the relation  $-I-\delta S_h\leq \delta S_c \leq- \delta S_h$. Given $\delta S_h<0$ (since $\langle q_h\rangle>0$),   when $I>-\delta S_h=\beta_h \langle q_h\rangle$, we can have  $0\leq-I-  \delta S_h\leq \delta S_c$, allowing net heat absorption from the cold reservoir $\langle q_c\rangle=-\delta S_c/\beta_c\geq0$.
(see Methods \ref{M1} ‘The relations between the {local} efficiency and the entropy production’ ).

According to Eq. (\ref{wm}),  the work fluctuations $\delta w^2=\langle w^2\rangle-\langle w\rangle^2$  can be determined
 \begin{align} \label{sdw}
\delta w^2=\delta w_{cor}^2+\epsilon(\delta w_{err}^2-\delta w_{cor}^2+\langle w_d\rangle^2)-\epsilon^2 \langle w_d\rangle^2,
\end{align} 
where  $\delta w_{cor}^2=\langle w_{cor}^2\rangle-\langle w_{cor}\rangle^2$  and  $\delta w_{err}^2=\langle w_{err}^2\rangle-\langle w_{err}\rangle^2$ represent the work fluctuations   under  correct and incorrect measurements, respectively.  In Eq. (\ref{sdw}), the quadratic term's coefficient is negative, and its symmetry axis is defined as $\epsilon_s=(\delta w_{err}^2-\delta w_{cor}^2+\langle w_d\rangle^2)/(2\langle w_d\rangle^2$). Based on these, if the control parameters $\lambda_{e,g}$ satisfy the conditions $\langle w_d\rangle<0$, and $\epsilon_s>\epsilon_0$,  the output work decreases  and the work fluctuations increase as  the  measurement error increases. 
However, when $\langle w_d\rangle>0$  and $\epsilon_s<\epsilon_0$ the output work increases  and the work fluctuations decrease as increasing measurement error. 

{Eqs (\ref{w}), (\ref{eff}), and (\ref{sdw}) give the general design principles for the information engines with optimal work, efficiency, and stability under given parameter conditions. (i) When the measurement error can be controlled, while the physical conditions of the engine, that is, the control parameters of the demon (thermal environment, energy level gap) cannot be changed, one can first calculate $\langle w_d\rangle$, $\epsilon_0,$ $\eta^{'}$, and $\epsilon_s$ based on the physical conditions, and then adjust the measurement error to enable the engine to operate at its optimal performance: for example, if $\langle w_d \rangle<0, \eta^{'}<0$, and $\epsilon_s>\epsilon_0$ ($\langle w_d \rangle>0, \eta^{'}>0$, and $\epsilon_s<\epsilon_0$), the measurement error need to be set at $\epsilon=0$ ($\epsilon=1$) for the optimal performance. (ii)When the measurement error cannot be changed, but the physical conditions can be altered, the physical conditions can be designed based on the magnitude of the measurement error to enable the engine to operate at high performance: for example, when the measurement error is small (large), the physical conditions should satisfy relation $\langle w_d \rangle<0, \eta^{'}<0$, and $\epsilon_s>\epsilon_0$  ($\langle w_d \rangle>0, \eta^{'}>0$, and $\epsilon_s<\epsilon_0$) for operating at high performance.
(iii) Eqs (\ref{w}), (\ref{eff}), and (\ref{sdw}) constitute a universal framework for predicting how any performances  of an information engine depends on measurement error. Once a control protocol is given, the conditional averages $\langle w_{cor}^m\rangle$, $\langle w_{err}^m\rangle$, $\langle q_{cor}\rangle$, and $\langle q_{err}\rangle$ can be computed, and the full error dependence of the work, efficiency, and fluctuations follows immediately. This decoupling of error dependence from trajectory-level thermodynamics is a practical tool for engine characterization and optimization, irrespective of which regime the engine operates in.
In a word,
 our approach offers distinct performance advantages over existing strategies. Enhancing measurement precision \cite{Sagawa20,Sag14} incurs significant energetic penalties, such as Landauer memory erasure costs and high-precision apparatus demands. Similarly, Bayesian filtering \cite{TK22, AT22, LC24, JD24} and delayed feedback \cite{KV25} demand complex computational overhead, and their net work output still vanishes under high noise. In contrast, our method completely bypasses these operational costs. By strategically configuring the demon's control parameters, our engine dose not expend energy fighting the error, and  can  operate robustly at any error level, uniquely utilizing measurement error to maximize performance.}

 \subsection{The bounds of the {local} efficiency}
Once the system interacts with the heat reservoirs, a correlation will also be established between the demon and the heat reservoirs \cite{Zhang22}. During the heat exchange process between the system and the heat reservoirs, not only will the correlation between the system and the demon change, but also the correlation between the demon and the heat reservoirs will change. The change of this correlation  can be quantified by the information on the flow of the demon to the heat reservoirs which is defined as $\delta I=\sum_{X_F,y}p(X_F,y)\{\ln[p(X_F|x_i,y)/p(X_B|x_f,y)]-\ln[p(X_F|x_i)/p(X_B|x_f)]\}$. Here,  $\ln[p(X_F|x_i,y)/p(X_B|x_f,y)]$ represents the entropy change of the reservoirs under the demon’s control \cite{sagawa12}, while  $\ln[p(X_F|x_i)/p(X_B|x_f)]$ corresponds to the coarse-grained entropy change of the reservoirs with  $X_B$ and  $x_f$ denoting the time-reversed trajectory of $X_F$ and the system's final state of a cycle. During a cycle, the total contribution of the demon to both the entropy of the system and the entropy of the heat reservoirs can be defined as the information dissipation $\Sigma_I=\sum_{X_F,y}p(X_F,y)\{\ln[p(X_F|y)/p(X_F)]-\ln[p(X_B|y)/p(X_B)]\}=I+\delta I$ \cite{Zeng21}. According to the fluctuation theorems of the  information dissipation $\langle \exp(-\Sigma_I)\rangle=1$, we have $\Sigma_I=I+\delta I\geq0$, which implies  $I \geq -\delta I$.  Substituting this inequality into Eq. (\ref{eta1}), one can obtain the lower bound of the TRIEs's {local} efficiency: 
\begin{eqnarray}\label{efflow}
\eta\geq\eta_{low}=\eta_C-\frac{ \delta I+ \Sigma_{tot}}{\beta_c\langle q_h\rangle}=\eta_C-\frac{\Sigma_I+\delta S_r}{\beta_c\langle q_h\rangle}.
\end{eqnarray}
Eq. (\ref{efflow}) indicates that increasing the thermodynamic {local}  efficiency of an information engine requires not only reducing the entropy change of the heat reservoirs, as is the case with traditional heat engines, but also minimizing information dissipation.

Furthermore, in reference \cite{Zeng21} the relation   $\Sigma_{tot}\geq\Sigma_I$ has been proven, indicating $\delta S_r\geq \delta I$.
Substituting this inequality into Eq. (\ref{eta1}),   the upper bound of the TRIEs's {local} efficiency can be obtained:
\begin{eqnarray}\label{etaup}
\eta&\leq&\eta_{up}=\eta_C-\frac{\delta I}{\beta_c \langle q_h\rangle}, 
\end{eqnarray}
The upper bound derived here is tighter than that obtained in Ref. \cite{Sagawa08}, namely $\eta_{up}\leq \eta_C+I/(\beta_c\langle q_h\rangle)$, owing to the relation $I\geq-\delta I$. Therefore, our conclusions does not violate the second law of thermodynamics.

 Combining Eqs. (\ref{efflow}) and (\ref{etaup}), the gap of the range of achievable efficiencies is  $\eta_{up}-\eta_{low}=\Sigma_{tot}/(\beta_c \langle q_h\rangle)$. This equation illustrates that a smaller total entropy production leads to a tighter and more precise assessment of {local} efficiency for the TRIEs.  When the entire cycle is reversible where $\Sigma_{tot}=0$, the {local} efficiency of the TRIEs is not less than the  Carnot efficiency $\eta=\eta_{up}=\eta_{low}=\eta_C+I/(\beta_c\langle q_h\rangle)$ which is the upper bound obtained by the Sagawa-ueda theorem \cite{Sagawa08}. At this point, the entropy change in the heat reservoirs is entirely caused by the information on the
flow of the demon to the heat reservoirs $\delta S_r=\delta I$.

\subsection{The global efficiency of the TRIEs}

{Even accounting for the cost of the demon  when defining efficiency, named the global efficiency,  the TRIEs can still defeat the Carnot efficiency.}

{If the energy required for the demon's measurement and memory resetting is provided by the hot reservoir and the  cold reservoir,  the global efficiency of the TRIEs cannot exceed the Carnot efficiency. The reason is as follows.
Consider the demon and the system as a single composite entity. Assume that during the measurement and erasure of the demon's memory, the demon exchanges heat only with the hot reservoir at inverse temperature $\beta_h$ and the cold reservoir at $\beta_c$, with the exchanged heat amounts denoted by $Q_h^{D}$ and $Q_c^{D}$, respectively.
After a cycle, the total entropy production of the composite system (system + demon+reservoirs) satisfies the relation
$\Sigma_{tot}=\delta S_r - \beta_h Q_h^{D} - \beta_c Q_c^{D}$, indicating $- \beta_h Q_h^{D} - \beta_c Q_c^{D}=I$.}
{Because any measurement and erasure protocol requires a net input of energy, we have $Q_h^{D} + Q_c^{D} \ge 0$, i.e., $Q_h^{D} \ge -Q_c^{D}$, leading to
 $-Q_c^{D} \ge I/(\beta_c - \beta_h)$.
Consequently,
$ Q_h^{D} \ge I/(\beta_c - \beta_h) \ge 0 \ge -I/(\beta_c - \beta_h) \ge Q_c^{D}.$
Therefore, if the demon  interacts with the two heat reservoirs, the heat that must flow from the hot reservoir to the demon is at least $Q_h^{D} = I/(\beta_c - \beta_h)$. If we incorporate this demonic dissipation into the definition of the efficiency, the total heat flow from the hot reservoir to the composite  entity is $\langle q_h \rangle + Q_h^{D}$, while the work delivered to the outside remains $\langle w \rangle$. The global efficiency of the composite entity then reads
$\langle w \rangle/(\langle q_h \rangle + Q_h^{D})$. Using inequality $\Sigma_{tot}\geq 0$, one can find $\langle w \rangle/(\langle q_h \rangle + Q_h^{D})\leq \eta_C$.
Hence, when both the hot and cold reservoirs are used to supply the energy for measurement and erasure, the global efficiency of the composite entity never exceeds the Carnot limit.}

{However, a different scenario arises if only the cold reservoir is used to provide the energy required for measurement and erasure. In  this case, according to the Landauer principle \cite{Landauer61},  the minimal required energy can be set as $Q_c^{D} = I/\beta_c$. At this time, the global efficiency of the composite entity should be defined as
\begin{equation}
    \eta_d = \frac{\langle w \rangle}{\langle q_h \rangle + Q_c^{D}}=\frac{\langle w \rangle}{\langle q_h \rangle + I/\beta_c}.
\end{equation}}
{Then, the condition for $\eta_d > \eta_C$ is
$-\delta S_r \ge \eta_C I$.
Since  $-\delta S_r\leq I$, and $0 \le \eta_C \le 1$, this inequality can  be satisfied. Thus, with cold reservoir providing energy,  the global efficiency of the TRIEs  can surpass the Carnot efficiency without violating any thermodynamic law.}

$
\eta_d = \frac{\langle w \rangle}{\langle q_h \rangle + Q_c^{D}}=\frac{\langle w \rangle}{\langle q_h \rangle + I/\beta_c}.
$

 \subsection{Example: Dual Otto cycles TRIE}
\subsubsection{Numerical Analysis}

In this section, we illustrate our findings by considering that  the controls exerted by the demon enable the system to operate as an Otto cycle, consisting of two adiabatic strokes and two isochoric strokes, as depicted in Fig. \ref{Model}(b). The detailed dynamics of the dual Otto cycles TRIE are described in the following section.

 Firstly, according to the measurement outcome, the demon  adiabatically  changes the system's  energy level gap from  $E_c$ to $E_y$ ($A\rightarrow B$). 
Throughout this adiabatic stroke, the system remains thermodynamically isolated from the heat reservoirs, ensuring that  the state of the system remains constant. This isolation leads  the work $w_{ch}$ produced by the system to be equal to the change in its internal energy. Subsequently, the demon controls the system to interact with the hot reservoir at an inverse temperature $\beta_h$ until  the Gibbs state is achieved ($B \rightarrow C$). During this isochoric heating stroke,  the system's energy level gap remains fixed at $E_y$. 
At the end of this stroke, the probabilities of the system occupying the excited and ground states conditioned on $y$ are given by $p(x_C=e|y)=\exp(-\beta_h E_y)/[1+\exp(-\beta_h E_y)]$, and  $p(x_C=g|y)=1/[1+\exp(-\beta_h E_y)]$, respectively.
 Since no work is performed during the isochoric stroke, the stochastic heat injection $q_h$ is equivalent to the increase of the system's internal energy. 
In the next step, the system is again isolated from the heat reservoirs, and the demon adiabatically tunes the system's energy level gap back to the initial value $E_c$ ($C\rightarrow D$).
As this process is adiabatic, the system’s state remains unchanged, and  the change in internal energy corresponds to the work $w_{hc}$ produced by the system. Finally, the demon controls the system to interact with the cold reservoir with fixed energy gap $E_c$, ultimately returning it to the initial thermal distribution ($D\rightarrow A$).
During this process no work is produced,   the absorbed heat $q_c$ equals the change in the system’s internal energy.  After completing a cycle, the total stochastic work  can be obtained by the relation $w=w_{ch}+w_{hc}$.

Assuming the populations for the system in the ground and excited states are 0 and 1, respectively, the average population changes for $y=e$ and $y=g$ conditioned on correct measurement during the hot isochoric stroke $B\rightarrow C$ can be expressed as  $\mathcal{N}_e=-p_ep(x_C=g|y=e)$ and $\mathcal{N}_g=p_gp(x_C=e|y=g)$, respectively. In addition,  the average population changes of the system    for $y=e$ and $y=g$ conditioned on  erroneous measurement are $\mathcal{M}_{e}=p_gp(x_C=e|y=e)$  and $\mathcal{M}_{g}=-p_ep(x_C=g|y=g)$, respectively. Then the average  work
$\langle w_{cor}\rangle$ and $\langle w_{err}\rangle$, the average heat $\langle q_{cor}\rangle$ and $\langle q_{err}\rangle$, and the average  of the square of the work $\langle w_{cor}^2\rangle$ and $\langle w_{err}^2\rangle$ can be expressed as  (see Appendix \ref{saw})
\begin{eqnarray}\label{hvjv}
   \langle w_{cor}\rangle
    &=&(E_e-E_c)N_e+(E_g-E_c)N_g,\nonumber\\
    \langle w_{err}\rangle&=&(E_e-E_c)M_e+(E_g-E_c)M_g,\nonumber\\
    \langle q_{cor} \rangle&=&E_eN_e+E_gN_g,
    \langle q_{err}\rangle 
    =E_eM_e+E_gM_g,\nonumber\\
    \langle w^2_{cor}\rangle&=&(E_g-E_c)^2N_g-(E_e-E_c)^2N_e,\nonumber\\
\langle w^2_{err}\rangle&=&(E_e-E_c)M_e-(E_g-E_c)^2M_g. 
\end{eqnarray}
Subsequently, the output work, the {local} efficiency, the work fluctuations,  and the heat $\langle q_c\rangle=\langle w\rangle-\langle q_h\rangle$ of the dual Otto cycles TRIE  can be determined. 

\begin{figure}[h]
\centering
 \begin{overpic}[width=0.9\textwidth]{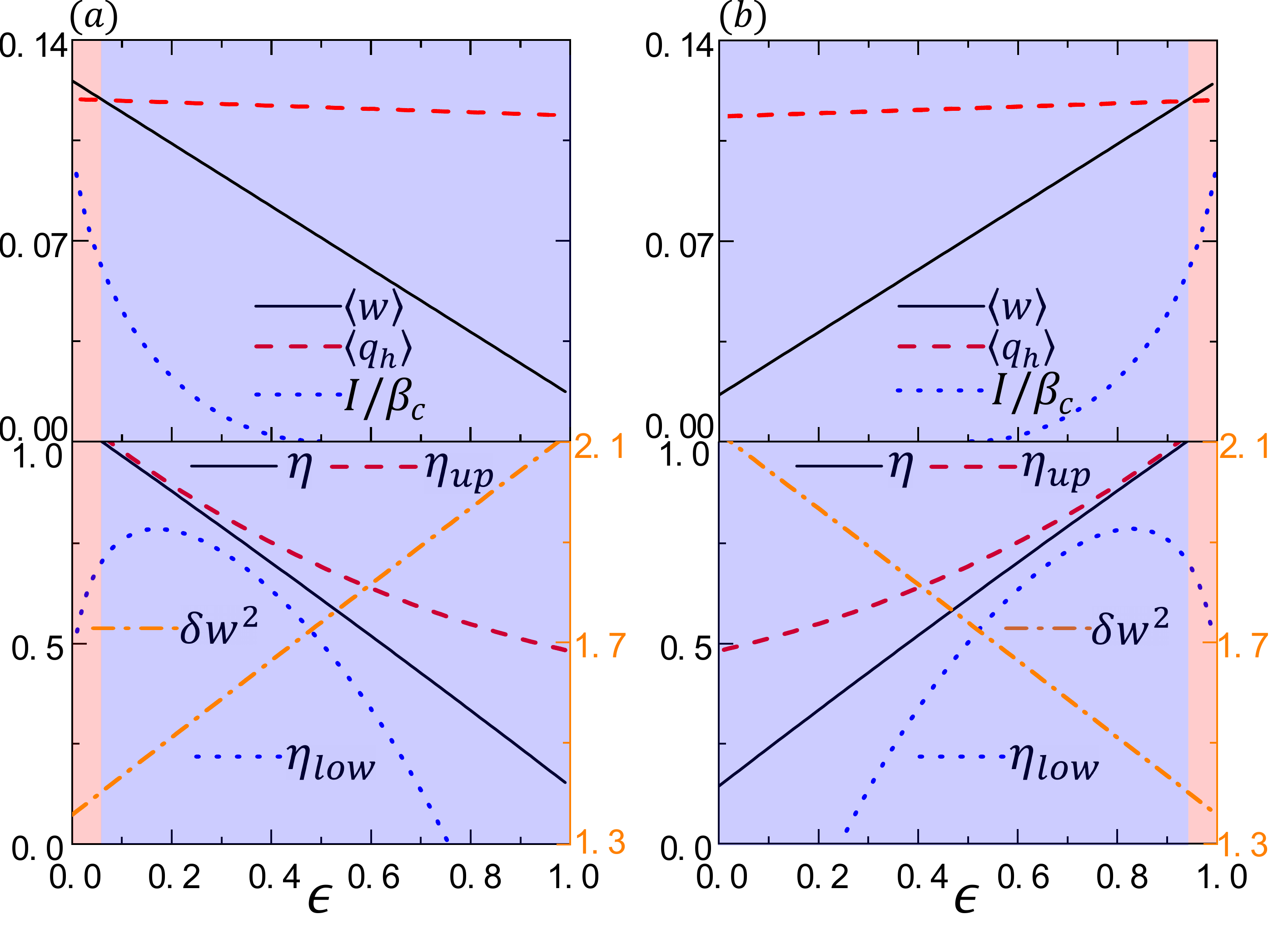}
\end{overpic}
\caption{The total output work $\langle w\rangle $, the heat $\langle q_{h}\rangle$, the work $I/\beta_c$ (upper panels),
   the {local} efficiency $\eta$,  and the work fluctuations (lower panels) as a functions of the measurement error $\epsilon$.  {The  blue,  and  red regions represent the regime where $0<\eta<1$, and $\eta=1$ respectively.}  In (a), $E_g=5.5$, $E_e=3$. In (b), $E_g=3$, $E_e=5.5$.
 The other parameters are  $E_c=1$, $\beta_c=2.5$, and $\beta_h=0.5$.}\label{effpe}
\end{figure}

In Fig.~\ref{effpe}(a), the work difference $\langle w_d\rangle$ is calculated as $\langle w_d\rangle =-0.11<0$. Thus, $\langle w\rangle$ decreases monotonically with increasing measurement error $\epsilon$. In this figure, the zero point of the work is $\epsilon_0=1.15>1$.  Consequently, the  dual Otto cycles TRIE can produce positive work under any measurement error. Furthermore, if the work difference   $\langle w_d\rangle$  in the SRIEs is also negative,
 the dual Otto cycles TRIE can produce more work than the SRIEs due to the maximum extractable work $I/\beta_c$ being smaller than the work $\langle w\rangle$.
Additionally, in Fig.~\ref{effpe}(a), the value of 
$\eta^{'}\langle q_h\rangle^2=-0.01$ indicates  $\eta^{'}<0$. {The corresponding measurement error for achieving perfect  efficiency is   $\epsilon_p = 0.06$. Thus, within the ranges $0\leq\epsilon<0.06$,  the dual Otto cycles TRIE can  achieve perfect local efficiency.}
 Finally,   owing to the symmetric axis $\epsilon_s=110.76>\epsilon_0$,  when $\epsilon=0$ in Fig. \ref{effpe}(a), the work, the {local} efficiency, and the stability of the dual Otto cycles TRIE can simultaneously reach their maximum values.

\begin{figure*}[h]
    \centering
    \includegraphics[width=0.6\linewidth]{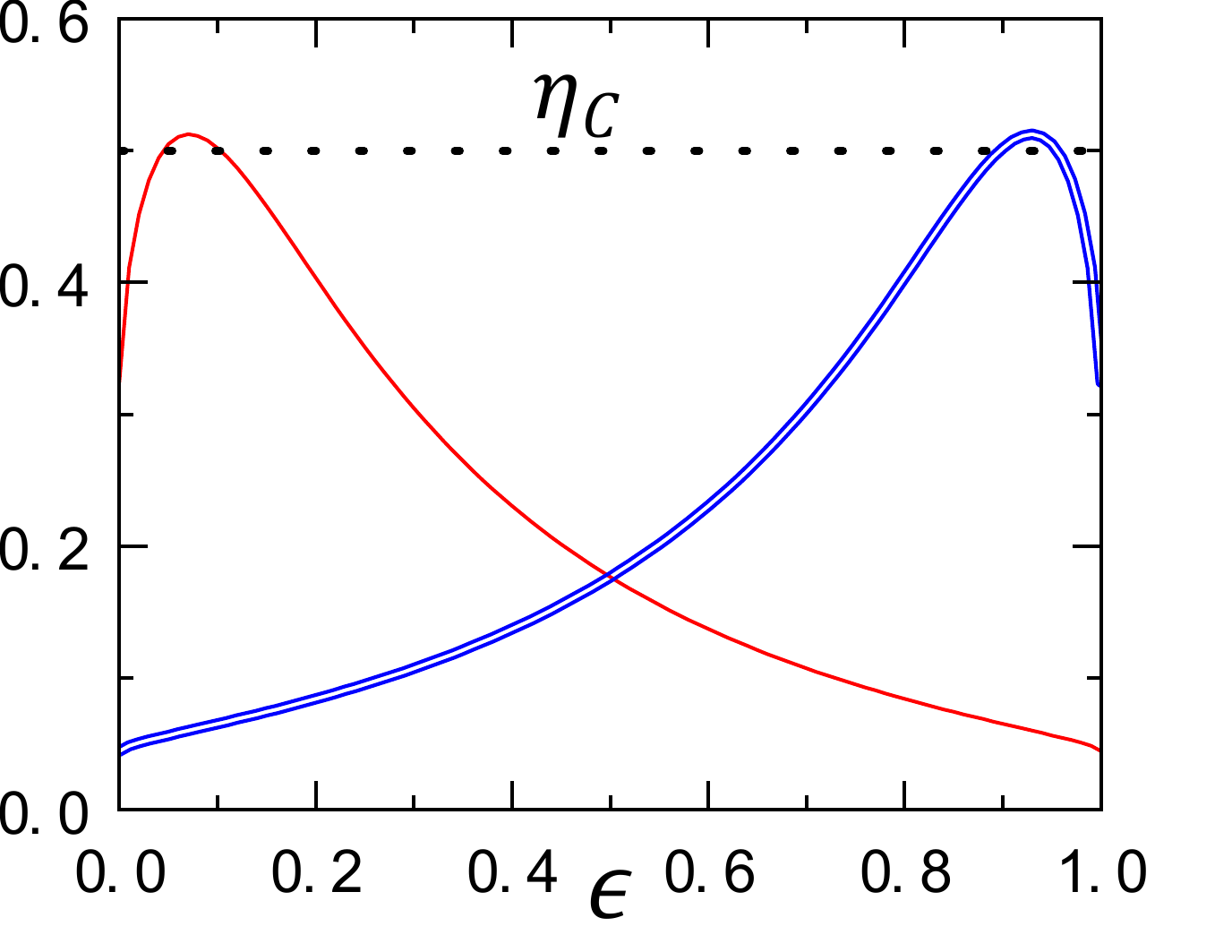}
    \caption{The global efficiency $\eta_d$ as a function of the measurement error. The parameters of the red solid curve are $E_g=10$, and  $E_e=6$. The parameters of the blue double solid curve are $E_g=6$, and  $E_e=10$. The other parameters are $E_c=4.5$ and $\beta_c=2\beta_h=1$.}
    \label{effd}
\end{figure*}

 In Fig. \ref{effpe}(b), since $\langle w_d\rangle=0.11>0$ and $\eta^{'}\langle q_h\rangle^2=0.01>0$, both the work and the {local} efficiency of the dual Otto cycles TRIE increase with measurement error, indicating that measurement error can be utilized to enhance the performance of information engines. In this figure, $\epsilon_0=-0.15<0$, $0<\epsilon_p= 0.94<1$, and $\epsilon_s=-109.76<\epsilon_0$. Therefore, the dual Otto cycles TRIE can produce positive work under any measurement error,  achieve perfect {local} efficiency, and optimize work, {local} efficiency, and stability simultaneously.
 
In Fig. \ref{effpe}, the black solid curve ($\eta$) is consistently positioned above the blue dotted curves ($\eta_{low}$) and below the red dashed curves ($\eta_{up}$), thereby confirming the validity of the bounds established by Eqs. (\ref{efflow}) and (\ref{etaup}). In addition, it also indicates that our conclusions do not violate the laws of thermodynamics.

{In Fig.~\ref{effd}, under both the error-suppression regime ($\langle w_d \rangle < 0$) and the error-exploitation regime ($\langle w_d \rangle > 0$), the global efficiency of the composite entity (system $+$ demon), denoted as $\eta_{d}$,  initially increases and then decreases with the measurement error $\epsilon$ increasing.
For the error-suppression regime, the reason is as follows. At small $\epsilon$, the mutual information $I$ acquired by the demon decreases rapidly with increasing measurement error, and the demon's dissipation  also drops quickly. Although the measurement error reduces the work output $\langle w \rangle$, it reduces the total heat flowing into the composite system  faster. Consequently, $\eta_{d}$ increases with $\epsilon$ in this initial stage. As the measurement error continues to grow, the mutual information gradually approaches zero, and the demon's dissipation becomes negligible compared to $\langle q_h\rangle$. In this intermediate stage, $\eta_{d}$ approaches the {local} efficiency $\eta$, and therefore decreases with further increase of $\epsilon$. When $\epsilon > 0.5$, the mutual information $I$ increases with $\epsilon$, and the demon's dissipation also rises, causing $\eta_{d}$ to decrease monotonically.}

{For the error-exploitation regime, the behavior is different. At small $\epsilon$, the demon's dissipation decreases while the work output increases monotonically with $\epsilon$, so $\eta_{d}$ initially rises. When $\epsilon$ exceeds $0.5$, the mutual information initially grows slowly with $\epsilon$, and $\eta_{d}$ remains close to $\eta$ who increases as the measurement error increases. As $\epsilon$ increases further, the information acquired by the demon grows rapidly, and the demon's dissipation also rises sharply. At this point, the heat flowing into the composite entity increases faster than the work output, causing $\eta_{d}$ to eventually decrease with $\epsilon$.}

{Importantly, Fig.~\ref{effd} demonstrates that the global efficiency $\eta_{d}$ can surpass the standard Carnot efficiency $\eta_C$ in both the error-suppression and error-exploitation regimes. This indicates that incorporating a demon into a conventional two-reservoir heat engine enables operation beyond the Carnot limit, provided the energy cost of  the demon is accounted for appropriately.} {However, the global efficiency of SRIEs cannot exceed 0.5, because the minimum energy injected from the surroundings into the composite system (the working system plus the demon) is $W + I/\beta_c$, indicating the SRIEs' global efficiency  $W/(W + I/\beta_c)<0.5.$}

\subsubsection{Experimental scheme}

The dual Otto cycles TRIE can be realized by using the electronic structure of a single trapped  $^{40}\mathrm{Ca}^{+}$. The spin degree of freedom can be  encoded in the $S_{1/2}$ and $S_{-1/2}$ energy levels, serving as the spin-up and spin-down states, respectively. The  work storage mechanism can be  implemented by the ion's motional vibrational modes. The total system Hamiltonian can be  decomposed into three  components \cite{Lutz24,OV23}: 
$H=H_{s}+H_{har}+H_{int}$. Here $H_{s}=\omega(t)\sigma_z/2+\Omega\sigma_x/2$ ($\hbar=1$) describes the system Hamiltonian, where $\omega(t)$  is a time-dependent control field  and $ \Omega$ represents the Rabi frequency of the driving field. $H_{har}=\omega_{har}a^{\dag}a$ represents the harmonic oscillator Hamiltonian for work storage,  with $\omega_{har}$ 
  representing the oscillator frequency 
 and $a^\dag$  
($a$) being the  creation (annihilation) operators.  The spin-oscillator interaction is governed by  $H_{int}=-(\tilde{\eta}\Omega/2)\sin(\omega_{har}t)\sigma_y$,  where   $\tilde{\eta}$ represents the Lamb-Dicke parameter.  $\sigma_{x,y,z}$ represent the Pauli operator.  The measurement 
 can be realized via two composite
light fields, with the  fluorescence frequency being the measurement outcome. The adiabatic strokes can be  achieved through slow modulation of $\omega$ \cite{JP29}. The isochoric strokes can be implemented by the modulation  of Rabi frequency which controls the transitioning rate between  the system's ground and excited state \cite{Yan22}.
The measurement process, adiabatic  stroke, and isochore stroke are detailed described in Methods \ref{M2} ‘The relations between the {local} efficiency and the entropy production’.

\subsection{Multi-level TRIEs}
Consider a system with $N+1$ discrete energy levels $x_i \in \{0, 1, \dots, N\}$. The demon performs a measurement on the system, yielding a measurement outcome $y \in \{0, 1, \dots, N\}$ with measurement accuracy characterized by
    $p(y = x_i | x_i) = 1 - \epsilon$.
The measurement error distribution is given by
$p(y \neq x_i | x_i) = a_{y x_i} \epsilon,$
where $a_{y x_i}$, being independent of measurement error, denotes the conditional probability of obtaining outcome  $y$ given state $x_i$, conditioned on erroneous measurement. These probabilities satisfy the conditions $0 \leq a_{y x_i} \leq 1$ and $\sum_{y=0}^{N} a_{y x_i}(1-\delta_{x_i,y}) = 1$ for all $x_i$. 
The probability of correct measurement is
$p(\mathcal{O}_{cor}) = \sum_{x_i} p(y=x_i | x_i) p(x_i) = 1 - \epsilon.$
The probability of erroneous measurement is
$p(\mathcal{O}_{err})=\sum_{x_i} p(x_i)\sum_{y} a_{y x_i}(1-\delta_{x_i,y}) \epsilon = \epsilon.$ Then, the average of the $m$-th power of the work conditioned on the correct and erroneous measurements can be expressed as:  $\langle w_{cor}^m\rangle=\sum_{X_F,y}p(X_F|x_i,y)p(y|x_i)p(x_i)w^m(X_F,y)\delta_{x_i,y}/p(\mathcal{O}_{cor})=\sum_{X_F}p(X_F|x_i,y=x_i)p(x_i)w^m(X_F)$, and $\langle w_{err}^m\rangle=\sum_{X_F,y}p(X_F|x_i,y)p(y|x_i)p(x_i)w^m(X_F,y)(1-\delta_{x_i,y})/p(\mathcal{O}_{err})=\sum_{X_F,y}p(X_F|x_i,y)a_{x_i,y}p(x_i)w^m(X_F,y)(1-\delta_{x_i,y})$. Thus, Eq. (\ref{wm}) can be obtained. Similarly, the
 heat $\langle q_h\rangle$ can still be expressed as $\langle q_h\rangle=(1-\epsilon)\langle q_{cor}\rangle+\epsilon \langle q_{err}\rangle$. Consequently,  we can still obtain the Eqs. (\ref{w}),
 (\ref{eff}), and (\ref{sdw}) for multi-level systems, 
indicating that our results are applicable to multi-level system. To substantiate this statement, we have designed  a three-level TRIE {and a $N$-level TRIE} in Appendix \ref{th} {and Appendix \ref{HO}}.

\section{Methods}
\subsection{The relations between the {local} efficiency and the entropy production}\label{M1}
 After the measurement, the probabilities for the demon's memory states  $y=e$ and $y=g$ can be expressed as:  $p(y=e)=p_e(1-\epsilon)+p_g\epsilon$ and $p(y=g)=p_e\epsilon+p_g(1-\epsilon)$, respectively. This measurement operation establishes a correlation between the system  and the demon. The correlation can be quantified by the mutual information \cite{TM60}, which reads
\begin{eqnarray}\label{IA}
I&=&\mathop{\sum}\limits_{x_i,y}p(y|x_i)p(x_i)\ln\frac{p(y|x_i)}{p(y)}=\mathop{\sum}\limits_{x_i}p(x_i)\mathop{\sum}\limits_{y}p(y|x_i)\ln p(y|x_i)-\mathop{\sum}\limits_{y}\ln p(y)\mathop{\sum}\limits_{x_i}p(x_i,y)\nonumber\\
&=&\mathop{\sum}\limits_{x_i}p(x_i)[\epsilon\ln\epsilon+(1-\epsilon)\ln(1-\epsilon)-\mathop{\sum}\limits_{y}p(y)\ln p(y)
=S_Y-S_\epsilon.
\end{eqnarray}
where $S_Y=-p(y=e)\ln p(y=e)-p(y=g)\ln p(y=g)$  denotes the entropy of the demon's memory after the measurement, and $S_\epsilon=-(1-\epsilon)\ln(1-\epsilon)-\epsilon\ln\epsilon$ represents the entropy associated with the measurement error.

We define the heat absorbed by the system from the hot and cold reservoirs as $\langle q_h\rangle$ and $\langle q_c\rangle$,  and assume that the heat capacity of the heat reservoir is very large, which indicates that the temperatures of the heat reservoirs remain constant during the heat exchange process.  In the context of Langevin or Markovian dynamics, the  entropy changes of the hot reservoir and the cold reservoir caused by heat exchanges $\langle q_h\rangle$ and $\langle q_c\rangle$ are given by \cite{Zeng21,Crook99,Jar00}
\begin{eqnarray}\label{shsc}
    \delta S_h=-\beta_h \langle q_h\rangle, \delta S_c=-\beta_c \langle q_c\rangle.
\end{eqnarray}
The negative sign comes from the regulation that the heat injected into the system is in the positive direction.
Correspondingly, the work done by the TRIEs can be expressed as
\begin{eqnarray}\label{w1}
  \langle w\rangle=-\delta S_h/\beta_h-\delta S_c/\beta_c.  
\end{eqnarray}

At the end of the one cycle, the  system returns to its initial thermal distribution. Consequently, the system and the demon become completely decoupled. Thus, upon completing one cycle, the change in mutual information is equal to the initial mutual information $I$.
Therefore, the total entropy production of the system, the demon, and the heat reservoirs after completing one cycle is given by:  $\Sigma_{tot}= \delta S_h+\delta S_c + I$.  In the context of information thermodynamics, the total entropy production $\Sigma_{tot}$ adheres to the relation \cite{sagawa12}:
\begin{equation}\label{neq}
\Sigma_{tot}=\delta S_h+\delta S_c+I\geq0.  
\end{equation}

Additionally, the TRIEs's {local} efficiency Eq. (\ref{eta1}) can be rewritten as:
\begin{eqnarray}\label{eff3}
    \eta=1+\frac{\beta_h}{\beta_c}\frac{\delta S_c}{\delta S_h}.
\end{eqnarray}
In the context of a heat engine, the condition $\langle q_h\rangle>0$ is  necessary requirement. This implies that the entropy change of the hot reservoir is negative, which can be mathematically represented as:  $\delta S_h=-\beta_h\langle q_h\rangle<0$. Consequently, the entropy change of the cold reservoir in a  heat engine  cycle (where the system returns to its initial distribution after one cycle) without the demon must satisfy the relation $\delta S_c\geq-\delta S_h>0$. This means that, in a standard heat engine cycle without a demon, the system is required to release heat to the cold reservoir, represented as $\langle q_c\rangle=-\delta S_c/\beta_c<0$. However, the TRIEs operate under different dynamics, where the system may not always release heat to the cold reservoir. Specifically, Eq. (\ref{neq}) can be reframed  as: $\delta S_c\geq -\delta S_h-I$ which may not always be greater than zero due to the relation $I\geq0$. If the condition $I>-\delta S_h$ is satisfied, meaning that the initial mutual information between the system and the demon exceeds the entropy flowing from the system to the hot reservoir, the relation $-\delta S_h-I\leq \delta S_c\leq0$ is permitted. Based on this analysis, in Eq. (\ref{eff3}), the TRIEs will achieve the perfect {local} efficiency $\eta=1$ if $\delta S_c=0$ is satisfied, accompanied by the positive work $\langle w\rangle=\langle q_h\rangle>0$.  In this scenario, the heat extracted from the hot reservoir can be completely converted into work, without any loss to the cold reservoir. At this stage, the role of the cold reservoir is not to absorb heat
 but to erase the correlation between the system and the demon. 
Furthermore, in the regime where $-\delta S_h-I\leq \delta S_c<0$, the system absorbed heat from the cold reservoir, $\langle q_c\rangle>0$. The maximum absorbs heat occurs at $\langle q_c\rangle=-(\delta S_h+I)/\beta_c$ when $\delta S_c$ reaches its minimum $\delta S_c=\delta S_h-I$. Within this regime, the system absorbs positive heat from both the cold and  hot reservoirs, and converts it into work. Consequently, the {local} efficiency of the TRIEs, defined as the ratio of work to heat input,  should be expressed as $\eta=\langle w\rangle/(\langle q_h\rangle+\langle q_c\rangle)=1$, which  is also the perfect {local} efficiency.

\subsection{The experimental scheme  for the dual Otto cycle TRIE}\label{M2}
In this section, we implement the dual Otto cycle TRIE using a single trapped $^{40}\mathrm{Ca}^+$  ion, where the $S_{1/2}$ and $S_{-1/2}$ represent the system’s excited and ground states and vibrational modes act as the energy battery. The total Hamiltonian is decomposed into system, harmonic oscillator, and interaction components \cite{Lutz24,OV23} $H=H_{s}+H_{har}+H_{int}$. The system Hamiltonian is defined as   $H_{s}=\omega(t)\sigma_z/2+\Omega\sigma_x/2$ (setting $\hbar=1$),  where $\omega(t)$ is a time-dependent driving field  and $\Omega$ is the Rabi frequency of the driving field. The work storage is governed by the harmonic oscillator Hamiltonian $H_{har}=\omega_{har}a^{\dag}a$, where $\omega_{har}$ is the oscillator frequency and $a^{\dag}$ ($a$)  is the creation (annihilation) operators. The spin-oscillator interaction follows
$H_{int}=-(\tilde{\eta}\Omega_{s}/2)\sin(\omega_{har}t)\sigma_y$, where $\tilde{\eta}$ the Lamb-Dicke parameter. Here,  $\sigma_{x,y,z}$ are the Pauli matrices. Regarding notation, the subscripts in $\sigma_{x,y}$ denote the spatial axes of the Pauli operators; otherwise, the variables $x$ and $y$ refer to the system state and the demon's measurement outcome, respectively.

\begin{figure}
\centering
\begin{overpic}[width=0.9\textwidth]{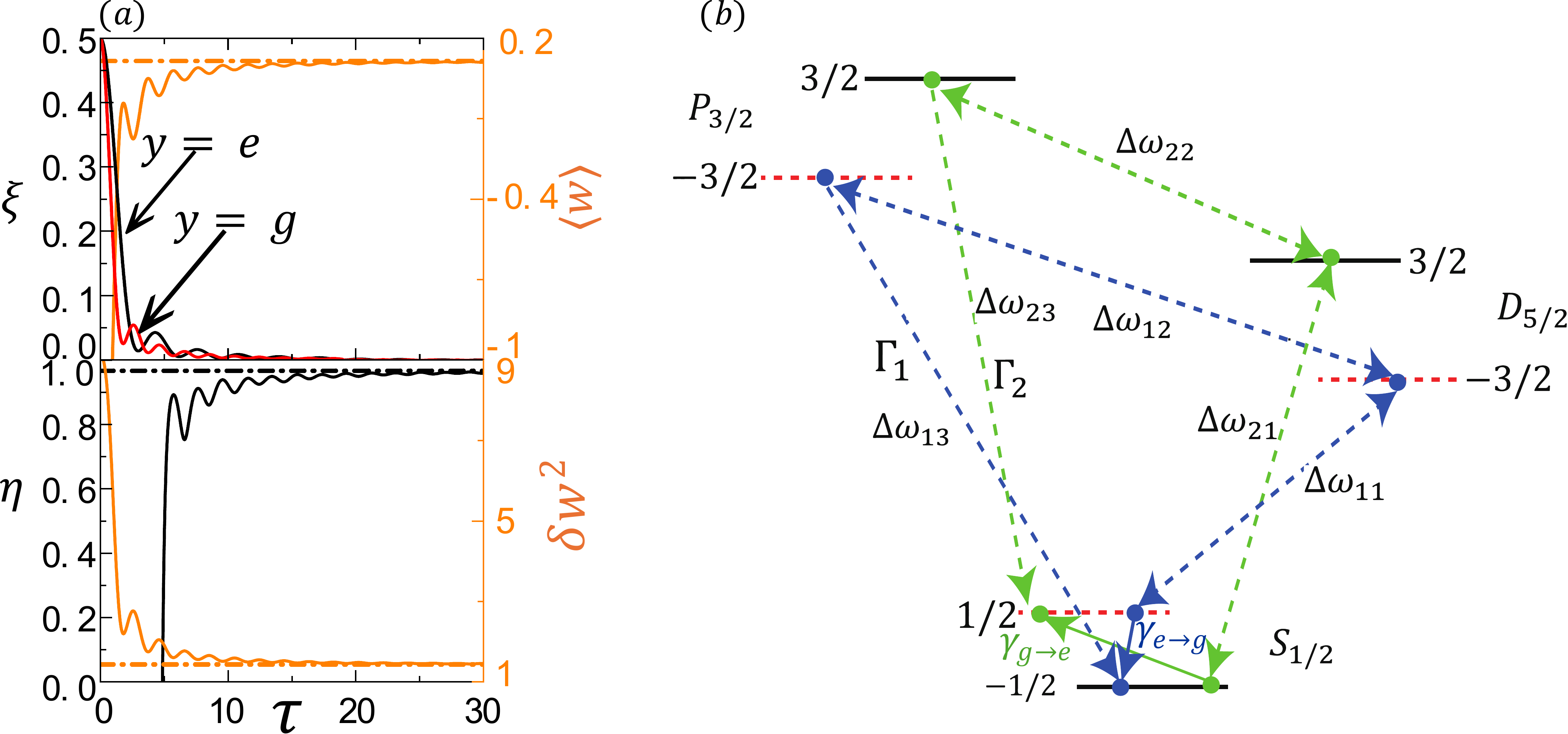}
\end{overpic}
    \caption{(a)The transition probability, the output work (upper panel), the {local} efficiency  and the work fluctuations (lower panel) as functions of the driving time $\tau$ with $\epsilon=0.1$. The solid  and dashed-dotted  curves  represent the simulated and theoretical values of the dual Otto cycles TRIS's  performance. The parameters are the same as Fig. \ref{effpe}(a). (b)Schematic illustration of $^{40}\mathrm{Ca}^{+}$ energy level and energy transitions \cite{Yan22}.}
    \label{ill}
\end{figure}

During the first adiabatic strokes, the system is driven over the time period  $\tau$ by the function $\omega_{ch}^{y}(t)=[\omega_c(1-t/\tau)+\omega_yt/\tau]\cos[\pi*t/(2\tau)]$. The  corresponding the Rabi frequency can be set to be $\Omega^y_{ch}=[\omega_c(1-t/\tau)+\omega_yt/\tau]\sin[\pi*t/(2\tau)]$.  Thus, in this process, the system Hamiltonian $H_{s}^y=\omega^{y}_{ch}\sigma_z/2+\Omega_{ch}^y\sigma_x/2$  does not commute with itself at different times, leading to nonadiabatic transitions. The probability of the system transitioning from state 
 $|i\rangle$ to state $|j\rangle$ is given by $\xi_{ch}^{y}=|\langle i|U_{ch}^y|j\rangle|^2$ \cite{{JP29}},  where unitary operator $U_{ch}^y=\mathcal{T}_{>}\exp[-i\int_0^{\tau}dtH_{s}^y(t)]$.  Here, $\mathcal{T}_{>}$ is the time-ordering operator. During the second adiabatic strokes, the system is driven over the same time period 
$\tau$ by 
 the functions $\omega_{hc}^{y}(t)=\omega_{ch}^y(\tau-t)$ and $\Omega_{hc}^{y}(t)=\Omega_{ch}^y(\tau-t)$.
 If the system is driven slowly, the condition 
  $\xi_{ch}^y=\xi_{hc}^y\rightarrow 0$, will be satisfied, representing the quantum adiabatic theorem, as shown in Fig. \ref{ill}(a).

The isochoric process is achieved through a laser-induced cascade process involving multiple electronic levels of the calcium ion. 
Specifically, for transitions from the excited (ground) state to the ground (excited) state, 
a laser with frequency $\Delta\omega_{11}^{k}$ ($\Delta\omega_{21}^{y}$ ) excites the electron from  $S_{1/2}$  ($S_{-1/2}$) to $D_{-3/2}$ ($D_{3/2}$) followed by another laser with frequency 
$\Delta\omega_{12}^{k}$  ($\Delta\omega_{22}^{k}$) driving the transition from $D_{-3/2}$ ($D_{3/2}$) 
  to  $P_{-3/2}$ ($P_{3/2}$),  where $\Delta\omega_{11}^{k}, \Delta\omega_{12}^{k}, \Delta\omega_{21}^{k}$ and $\Delta\omega_{22}^{k}$, correspond to the energy differences between $S_{1/2}$ and $D_{-3/2}$, $D_{-3/2}$ and $P_{-3/2}$, $S_{-1/2}$ and $D_{3/2}$, and $D_{3/2}$ and $P_{3/2}$.  Here $k=c,e,g$. The electron then decays radiatively to  $S_{-1/2}$ ($S_{1/2}$) with decay rate $\Gamma_1$ ($\Gamma_2$) as shown in Fig. \ref{ill}(b). The effective  transition rates from $S_{1/2}$  to $S_{-1/2}$  and  from $S_{-1/2}$ to $S_{1/2}$  after these transitions are given by \cite{Yan22}
  \begin{eqnarray}
      \gamma_{e\rightarrow g}^{k}=(\Omega_{11}^{k})^2\Gamma_1/(\Omega_{12}^{k})^2, \gamma_{g\rightarrow e}=(\Omega_{21}^{k})^2\Gamma_2/(\Omega_{22}^{k})^2,
  \end{eqnarray}
 where $\Omega_{11}^{k}, \Omega_{12}^{k}, \Omega_{21}^{k}$,  and $\Omega_{22}^{k}$ are the Rabi frequencies corresponding to the laser frequencies $\Delta\omega_{11}^{k}, \Delta\omega_{12}^{k}, \Delta\omega_{21}^{k}$ and $\Delta\omega_{22}^{k}$,  respectively. To ensure thermal equilibrium, the transition rates satisfy the detailed balance condition: $\gamma_{e\rightarrow g}^{c}/\gamma_{g\rightarrow e}^{c}=\exp(\beta_{c}\omega_c)$,  
 $\gamma_{e\rightarrow g}^{e}/\gamma_{g\rightarrow e}^{e}=\exp(\beta_{h}\omega_e)$, and $\gamma_{e\rightarrow g}^{g}/\gamma_{g\rightarrow e}^{g}=\exp(\beta_{h}\omega_g)$.

The measurement  process is simulated by two composite light fields, defined as: 
\begin{eqnarray}\label{EE}
     E_e(t)&=&(1-\epsilon)\cos(\Delta\omega_{11}^{c}t)+\epsilon\cos(\Delta\omega_{23}^{c}t), \nonumber\\
E_g(t)&=&\epsilon\cos(\Delta\omega_{21}^{c}t)+(1-\epsilon)\cos(\Delta\omega_{13}^{c}t).
\end{eqnarray}
Here, the frequency differences
$\Delta\omega_{11}^{c}$,  $\Delta\omega_{23}^{c}$,
$\Delta\omega_{21}^{c}$, and $\Delta\omega_{13}^{c}$,  correspond to the energy differences between the following electronic levels:  $S_{1/2}$ and $D_{-3/2}$, $S_{1/2}$ and $P_{3/2}$,
$S_{-1/2}$ and $D_{3/2}$,  and
 $S_{-1/2}$ and $P_{-3/2}$ at the initial point $A$, respectively. 
The composite light fields $E_e(t)$  or $E_g(t)$   are used to measure whether the electron is in the excited  or ground state through induced transitions and subsequent fluorescence measurements.  The four possible photon frequencies after injecting these composite light fields provide distinct measurement outcomes:
(a) Fluorescence at $\Delta\omega_{11}^c$ indicates a correct measurement of the excited state;  (b) Fluorescence at $\Delta\omega_{23}^c$ indicates an incorrect measurement of the excited state;  (c) Fluorescence at $\Delta\omega_{21}^c$ indicates an incorrect measurement of the ground state;
(d) Fluorescence at $\Delta\omega_{13}^c$ indicates a correct measurement of the ground state. 

To obtain the distributions of the   work and heat,   the fluorescence frequency, the population change of the harmonic oscillator during the hot isochoric stroke, and the population change of the harmonic oscillator upon cycle completion should be recorded during each cycle. The stochastic heat $q_h$ can be determined by combining the fluorescence frequency with the population change during the hot isochoric stroke. Similarly, the stochastic work $w$ can be determined by combining the fluorescence frequency with the population change at the end of the cycle.  The occurrences of each heat $q_h$ and work $w$  are then counted. These counts are normalized by dividing by the total number of cycles, yielding the probability distributions for heat and work. The performance parameters of the dual Otto cycle TRIE can subsequently be evaluated using these distributions. 

In Fig. (\ref{ill})(a), we have simulated our experimental plan, with dashed-dotted  and solid curves representing theoretical and simulated values. From this Figure, we can see that as the driving time increases, the transition probability in the adiabatic process tends to zero, and the simulated value tends to the theoretical value. Notice, the experimental plan above was not implemented in the laboratory.

{Furthermore, a closely related experimental realization has recently been reported by Zhang et al. \cite{Zhang2025}, where a single trapped $^{171}\mathrm{Yb}^+$ ion is used as an information engine that incorporates measurement, adiabatic, and thermalizing strokes. These  strokes are the same essential steps required for the dual Otto cycles TRIE proposed here, further indicating that our results are easily achievable experimentally.}

\section{Discussions}
In this work, we investigate  a class  of two-reservoir information engines (TRIEs) with two-level working system.  We show that the output work, the thermodynamics {local} efficiency,  and stability  of the TRIEs are the  monotonic functions of the measurement error. This result guides us how to adjust external parameters to enable the TRIEs to produce positive work under any measurement error, {and}  provide  the parameter regimes for the TRIEs optimizing these performance parameters by the measurement error. This converts measurement error from an unfavorable factor to an available resource.
{Moreover, with the demon's energy cost provided entirely by the cold reservoir, the TRIEs' efficiency (demon cost included) can still exceed the Carnot limit.}
Finally, to further validate our results, we propose a dual Otto cycles TRIE as a  example and outline an experimental scheme for  its
implementation  based on a trapped $^{40}\mathrm{Ca}^+$ ion. Notably,  our results can be applied to the TRIEs with multi-level system.  These conclusions demonstrate that TRIEs have better robustness than SRIEs,   can have higher  efficiency than traditional heat engines, and can utilize measurement error to optimize information engines. These advantages enhance the practicality of information engines in reality. {Especially,  we emphasize that our results are applied to the TRIEs possessing a discrete-level working systm, and a demon with uniform measurement accuracy. Furthermore, it remains an open question whether analogous error-exploitation mechanisms can be extended to continuous systems (such as Brownian particles) or to protocols with state-dependent measurement noise. By demonstrating how discrete feedback protocols can be engineered to bypass costly error-suppression overheads, our work provides a practical design strategy for noise-resilient micro- and quantum thermal machines, with potential implications for microscopic energy harvesting and quantum control.}

\bmhead{Acknowledgements}
Y. Xiao thanks the support in part by  the National Nature Science Foundation of China Grants No.12234019.
\bmhead{Data availability}The data that support the findings of this study are available from the corresponding author on request.
\bmhead{Code availability}The  code that supports the findings of this study is available from the corresponding author on request.
\bmhead{Author contributions}Y. X.,  Q. Z. and J. W. conceived the idea. Q. Z. and J. W. supervised the study.  Y. X.,  Q. Z. and J. W. developed the theoretical model. Y. X. performed the numerical simulations and  designed the experimental proposal. Q. Z. and J. W. contributed critically to the core conceptual  framework and physical explanations  of the results.
All authors discussed the results and wrote the final manuscript.

\begin{appendices}
\section{The performance of the dual Otto cycles TRIE}\label{saw}
According to the cycle of the dual Otto cycles TRIE, we have compiled Tab. \ref{tab1}, which presents the work, heat exchanges, and probabilities of occurrence for each stochastic trajectory. From Tab. \ref{tab1},  the average  work
$\langle w_{cor}\rangle$ and $\langle w_{err}\rangle$, the average heat $\langle q_{cor}\rangle$ and $\langle q_{err}\rangle$, and the average  of the square of the work $\langle w_{cor}^2\rangle$ and $\langle w_{err}^2\rangle$ can be determined which are as follows:

\begin{table}[h]
\caption{The work, the heat exchanges and probabilities  of the  stochastic trajectory $X_F$ conditioned on the initial state $x_i$ and the measurement outcome $y$ in the dual Otto cycles TRIE are presented.}\label{tab1}%
\begin{tabular*}{\textwidth}
{@{\extracolsep\fill}ccccccccc}
\toprule
$y$ & $x_i$  & $x_C$  & $x_f$ & $w_{ch}$ & $q_h$ & $w_{hc}$& $q_c$ & $p(X_F|x_i,y)$\\
\hline
$g$ & $e$   & $e$ & $e$ & $E_c-E_g$ & $0$ & $E_g-E_c$ & $0$ & $ p(x_C=e|y=g)p_e$\\

$g$ & $e$  & $e$  & $g$ & $E_c-E_g$ & $0$ & $E_g-E_c$ & $-E_c$ & $ p(x_C=e|y=g)p_g$  \\

$g$ & $e$  & $g$  & $e$ & $E_c-E_g$ & $-E_g$ & $0$ & $E_c$ & $p(x_C=g|y=g)p_e$ \\

$g$ & $e$ &   $g$  & $g$ & $E_c-E_g$ & $-E_g$ & $0$ & $0$ & $p(x_C=g|y=g)p_g$\\

$e$ & $e$ &  $e$  & $e$ & $E_c-E_e$ & 0 & $E_e-E_c$ & $0$ & $p(x_c=e|y=e)p_e$\\

$e$ & $e$ &  $e$  & $g$ & $E_c-E_e$ & 0 & $E_e-E_c$ & $-E_c$ & $p(x_c=e|y=e)p_g$\\

$e$ & $e$  & $g$  & $e$ & $E_c-E_e$ & $-E_e$ & $0$ & $E_c$ & $p(x_C=g|y=e)p_e$\\

$e$ & $e$ & $g$  & $g$ & $E_c-E_e$ & $-E_e$ & $0$ & $0$ & $p(x_C=g|y=e)p_g$\\

$g$ & $g$ & $e$  & $e$ & $0$ & $E_g$ & $E_g-E_c$ & $0$ & $p(x_C=e|y=g)p_e$\\

$g$ & $g$  & $e$  & $g$ & $0$ & $E_g$ & $E_g-E_c$ & $-E_c$ & $p(x_C=e|y=g)p_g$\\

$g$ & $g$  & $g$  & $e$ & $0$ & $0$ & $0$ & $E_c$ & $p(x_C=g|y=g)p_e$ \\

$g$ & $g$  & $g$  & $g$ & $0$ & $0$ & $0$ & $0$ & $p(x_C=g|y=g)p_g$\\

$e$ & $g$  & $e$ & $e$ & $0$ & $E_e$ & $E_e-E_c$ & $0$ & $p(x_C=e|y=e)p_e$\\

$e$ & $g$  & $e$  & $g$ & $0$ & $E_e$ & $E_e-E_c$ & $-E_c$ & $p(x_C=e|y=e)p_g$\\

$e$ & $g$  & $g$  & $e$ & $0$ & $0$ & $0$ & $E_c$ & $p(x_C=g|y=g)p_e$\\

$e$ & $g$ & $g$  & $g$ & $0$ & $0$ & $0$ & $0$ & $p(x_C=g|y=g)p_e$\\
\botrule
\end{tabular*}
\end{table}

\newpage

\begin{eqnarray}
   \langle w_{cor}\rangle&=&\frac{1}{1-\epsilon}\sum_{X_F,y}p(X_F|x_i,y)p(y|x_i)p(x_i)w(X_F,y)\delta_{x_i,y}\nonumber\\
    &=&\frac{1}{1-\epsilon}[-(E_e-E_c)p(x_C=g|y=e)p(y=e|x_i=e)p_e\nonumber\\
    &+&(E_g-E_c)p(x_C=e|y=g)p(y=g|x_i=g)p_g]\nonumber\\
    &=&-(E_e-E_c)p(x_C=g|y=e)p_e+(E_g-E_c)p(x_C=e|y=g)p_g\nonumber\\
    &=&(E_e-E_c)N_e+(E_g-E_c)N_g,\label{swcor}\\
    \langle w_{err}\rangle&=&\frac{1}{\epsilon}\sum_{X_F,y}p(X_F|x_i,y)p(y|x_i)p(x_i)w(X_F,y)(1-\delta_{x_i,y})\nonumber\\
    &=&\frac{1}{\epsilon}[(E_e-E_c)p(x_C=e|y=e)p(y=e|x_i=g)p_g\nonumber\\
    &-&(E_g-E_c)p(x_C=g|y=g)p(y=g|x_i=e)p_e]\nonumber\\
    &=&(E_e-E_c)p(x_C=e|y=e)p_g-(E_g-E_c)p(x_C=g|y=g)p_e\nonumber\\
    &=&(E_e-E_c)M_e+(E_g-E_c)M_g,\label{swerr}
   \end{eqnarray}

\newpage
    \begin{eqnarray}  
     \langle q_{cor} \rangle&=&\frac{1}{1-\epsilon}\sum_{X_F,y}p(X_F|x_i,y)p(y|x_i)p(x_i)q_h(X_F,y)\delta_{x_i,y}\nonumber\\
    &=&\frac{1}{1-\epsilon}[-E_ep(x_C=g|y=e)p(y=e|x_i=e)p_e\nonumber\\
    &+&E_gp(x_C=e|y=g)p(y=g|x_i=g)p_g]\nonumber\\
    &=&-E_ep(x_C=g|y=e)p_e+E_gp(x_C=e|y=g)p_g=E_eN_e+E_gN_g,\label{sqcor}\\
    \langle q_{err}\rangle&=& \frac{1}{\epsilon}\sum_{X_F,y}p(X_F|x_i,y)p(y|x_i)p(x_i)q_h(X_F,y)(1-\delta_{x_i,y})\nonumber\\
    &=&\frac{1}{\epsilon}[E_ep(x_C=e|y=e)p(y=e|x_i=g)p_g\nonumber\\
    &-&E_gp(x_C=g|y=g)p(y=g|x_i=e)p_e]\nonumber\\
    &=&E_ep(x_C=e|y=e)p_g-E_gp(x_C=g|y=g)p_e
    =E_eM_e+E_gM_g. \label{sqerr}\\
    \langle w^2_{cor}\rangle&=&\frac{1}{1-\epsilon}\sum_{X_F,y}p(X_F|x_i,y)p(y|x_i)p(x_i)w^2(X_F,y)\delta_{x_i,y}\nonumber\\
    &=&\frac{1}{1-\epsilon}[(E_e-E_c)^2p(x_C=g|y=e)p(y=e|x_i=e)p_e\nonumber\\
    &+&(E_g-E_c)^2p(x_C=e|y=g)p(y=g|x_i=g)p_g]\nonumber\\
    &=&(E_e-E_c)^2p(x_C=g|y=e)p_e+(E_g-E_c)^2p(x_C=e|y=g)p_g\nonumber\\
    &=&(E_g-E_c)^2N_g-(E_e-E_c)^2N_e,\label{swcor2}\\
\langle w^2_{err}\rangle&=&\frac{1}{\epsilon}\sum_{X_F,y}p(X_F|x_i,y)p(y|x_i)p(x_i)w^2(X_F,y)(1-\delta_{x_i,y})\nonumber\\
   &=&\frac{1}{\epsilon}[(E_e-E_c)^2p(x_C=e|y=e)p(y=e|x_i=g)p_g\nonumber\\
   &+&(E_g-E_c)^2p(x_C=g|y=g)p(y=g|x_i=e)p_e]\nonumber\\
    &=&(E_e-E_c)^2p(x_C=e|y=e)p_g+(E_g-E_c)^2p(x_C=g|y=g)p_e\nonumber\\
    &=&(E_e-E_c)M_e-(E_g-E_c)^2M_g,\label{swerr2}
\end{eqnarray}
Here, $N_e=-p_ep(x_C=g|y=e)$, $N_g=p_gp(x_C=e|y=g)$  represent the changes of the system population when the measurements are free-error for $y=e$ and $y=g$ during the hot isochoric stroke. The terms    $M_e=p_gp(x_C=e|y=e)$ and  $M_g=-p_ep(x_C=g|y=g)$ are the changes of the system population  when the measurements  are fully error for $y=e$ and $y=g$ during the hot isochoric stroke.

 \section{The dual Otto cycle three-level TRIE}\label{th}
 In this section, we design an dual Otto cycles  three-level TRIE. The system is initially prepared in thermal equilibrium with a cold reservoir at inverse temperature $\beta_c$. 
The occupation probabilities  for the ground state ($g$), first excited state ($e_1$), and second excited state ($e_2$) are given by:
    $p_g = 1/[1 + \exp(-\beta_c E_c) + \exp(-2\beta_c E_c)], 
    p_{e_1} = \exp(-\beta_c E_c)/[1 + \exp(-\beta_c E_c) + \exp(-2\beta_c E_c)],$ and $ 
    p_{e_2} =\exp(-2\beta_c E_c)/[1 + \exp(-\beta_c E_c) + \exp(-2\beta_c E_c)],$
where the corresponding energy levels are $0$, $E_c$, and $2E_c$, respectively.

Subsequently, the demon performs a measurement on the system state $x_i$ ($x_i \in \{g, e_1, e_2\}$), obtaining outcome $y$ ($y \in \{g, e_1, e_2\}$). The accuracy of the demon's measurement is quantified by
    $p(y = x_i | x_i) = 1-\epsilon$,
where $\epsilon$  represents the measurement error. We assume that this error is uniformly distributed among the two remaining energy levels, such that
    $p(y \neq x_i | x_i) = \epsilon/2$.
Using the method in Eq. (\ref{IA}), after such a measurement, the information $I$ obtained by the demon from the three-level system can be determined.

The thermodynamic cycle of this three-level system follows a protocol analogous to that of the dual Otto cycles TRIE. Based on the measurement outcome $y$,the demon first controls the system through an adiabatic compression stroke, during which the energy gap between adjacent levels is increased from  $E_c$ to $E_y$. Following this, with the energy gap fixed at  $E_y$,  the system is thermally coupled to a hot reservoir at an inverse temperature $\beta_h$ and eventually reaches thermal equilibrium. At this equilibrium state, the system's probability distributions are expressed as follows
$ p(x_C = g | y) = 1/[1 + \exp(-\beta_h E_y) + \exp(-2\beta_h E_y)], 
    p(x_C = e_1 | y) = \exp(-\beta_h E_y)/[1 + \exp(-\beta_h E_y) + \exp(-2\beta_h E_y)], 
    p(x_C = e_2 | y) = \exp(-2\beta_h E_y)/[1 + \exp(-\beta_h E_y) + \exp(-2\beta_h E_y)]$.
   Subsequently, the system undergoes an adiabatic expansion stroke, whereby the energy gap returns to its initial value $E_c$.  Finally, the system is coupled to the cold reservoir, reaching thermal equilibrium and thus returning to the initial probability distribution, thereby completing the thermodynamic cycle.

The dual Otto cycles three-level TRIE has a total of $3^4$ distinct stochastic trajectories. Due to the considerable complexity of the trajectory data, the stochastic trajectory will no longer be displayed. After completing the one cycle, the work $\langle w_{cor}\rangle$ and $\langle w_{err}\rangle$, the heat $\langle q_{cor}\rangle$ and  $\langle q_{cor}\rangle$, and  the average of the squared work $\langle w_{cor}^2\rangle$ and $\langle w_{err}^2\rangle$    are calculated as
\begin{eqnarray}
   \langle w_{cor}\rangle&=&\frac{1}{1-\epsilon}\sum_{X_F,y}p(X_F|x_i,y)p(y|x_i)p(x_i)w(X_F,y)\delta_{x_i,y}\nonumber\\
    &=&(E_g-E_c)[p(x_C=e_1|y=g)p_g+2p(x_C=e_2|y=g)p_g]\nonumber\\
    &+&(E_{e_1}-E_c)[p(x_C=e_2|y=e_1)p_{e_1}-p(x_C=g|y=e_1)p_{e_1}]\nonumber\\
    &-&(E_{e_2}-E_c)[2p(x_C=g|y=e_2)p_{e_2}+p(x_C=e_1|y=e_2)p_{e_2}],
    \end{eqnarray}
    \newpage
    \begin{eqnarray}
    \langle w_{err}\rangle&=&\frac{1}{\epsilon}\sum_{X_F,y}p(X_F|x_i,y)p(y|x_i)p(x_i)w(X_F,y)(1-\delta_{x_i,y})\nonumber\\
&=&\frac{(E_g-E_c)}{2}\{[p(x_C=e_2|y=g)-p(x_C=g|y=g)]p_{e_1}\nonumber\\
&-&[2p(x_C=g|y=g)+p(x_C=e_1|y=g)]p_{e_2}\}\nonumber\\
    &+&\frac{(E_{e_1}-E_c)}{2}\{[p(x_C=e_1|y=e_1)+2p(x_C=e_2|y=e_1]p_g\nonumber\\
    &-&[2p(x_C=g|y=e_1)+p(x_C=e_1|y=e_1)]p_{e_2}\}
    \nonumber\\
     &+&\frac{(E_{e_2}-E_c)}{2}\{[p(x_C=e_1|y=e_2)+2p(x_C=e_2|y=e_2)]p_g\nonumber\\
     &+&[p(x_C=e_2|y=e_2)-p(x_C=g|y=e_2)]p_{e_1}\},\\
   \langle q_{cor}\rangle&=&\frac{1}{1-\epsilon}\sum_{X_F,y}p(X_F|x_i,y)p(y|x_i)p(x_i)q_h(X_F,y)\delta_{x_i,y}\nonumber\\
&=&E_g[p(x_C=e_1|y=g)p_g+2p(x_C=e_2|y=g)p_g]\nonumber\\
&+&E_{e_1}[p(x_C=e_2|y=e_1)p_{e_1}-p(x_C=g|y=e_1)p_{e_1}]\nonumber\\
    &-&E_{e_2}[2p(x_C=g|y=e_2)p_{e_2}+p(x_C=e_1|y=e_2)p_{e_2}],\\
    \langle q_{err}\rangle&=&\frac{1}{\epsilon}\sum_{X_F,y}p(X_F|x_i,y)p(y|x_i)p(x_i)w(X_F,y)(1-\delta_{x_i,y})\nonumber\\
&=&\frac{E_g}{2}\{[p(x_C=e_2|y=g)-p(x_C=g|y=g)]p_{e_1}\nonumber\\
&-&[2p(x_C=g|y=g)+p(x_C=e_1|y=g)]p_{e_2}\}\nonumber\\
    &+&\frac{E_{e_1}}{2}\{[p(x_C=e_1|y=e_1)+2p(x_C=e_2|y=e_1]p_g\nonumber\\
    &-&[2p(x_C=g|y=e_1)+p(x_C=e_1|y=e_1)]p_{e_2}\}
    \nonumber\\
     &+&\frac{E_{e_2}}{2}\{[p(x_C=e_1|y=e_2)+2p(x_C=e_2|y=e_2)]p_g\nonumber\\
     &+&[p(x_C=e_2|y=e_2)-p(x_C=g|y=e_2)]p_{e_1}\},\\
   \langle w_{cor}^2\rangle&=&\frac{1}{1-\epsilon}\sum_{X_F,y}p(X_F|x_i,y)p(y|x_i)p(x_i)w^2(X_F,y)\delta_{x_i,y}\nonumber\\
    &=&(E_g-E_c)^2[p(x_C=e_1|y=g)p_g+4p(x_C=e_2|y=g)p_g]\nonumber\\
    &+&(E_{e_1}-E_c)^2[p(x_C=e_2|y=e_1)p_{e_1}+p(x_C=g|y=e_1)p_{e_1}]\nonumber\\
    &+&(E_{e_2}-E_c)^2[4p(x_C=g|y=e_2)p_{e_2}+p(x_C=e_1|y=e_2)p_{e_2}],
\end{eqnarray}
\newpage
\begin{eqnarray}
    \langle w_{err}^2\rangle&=&\frac{1}{\epsilon}\sum_{X_F,y}p(X_F|x_i,y)p(y|x_i)p(x_i)w^2(X_F,y)(1-\delta_{x_i,y})\nonumber\\
&=&\frac{(E_g-E_c)^2}{2}\{[p(x_C=e_2|y=g)+p(x_C=g|y=g)]p_{e_1}\nonumber\\
&+&[4p(x_C=g|y=g)+p(x_C=e_1|y=g)]p_{e_2}\}\nonumber\\
    &+&\frac{(E_{e_1}-E_c)^2}{2}\{[p(x_C=e_1|y=e_1)+4p(x_C=e_2|y=e_1]p_g\nonumber\\
    &+&[4p(x_C=g|y=e_1)+p(x_C=e_1|y=e_1)]p_{e_2}\}
    \nonumber\\
     &+&\frac{(E_{e_2}-E_c)^2}{2}\{[p(x_C=e_1|y=e_2)+4p(x_C=e_2|y=e_2)]p_g\nonumber\\
     &+&[p(x_C=e_2|y=e_2)+p(x_C=g|y=e_2)]p_{e_1}\}.
     \end{eqnarray}

 \begin{figure*}[h]
 \centering
 \begin{overpic}[width=0.9\textwidth]{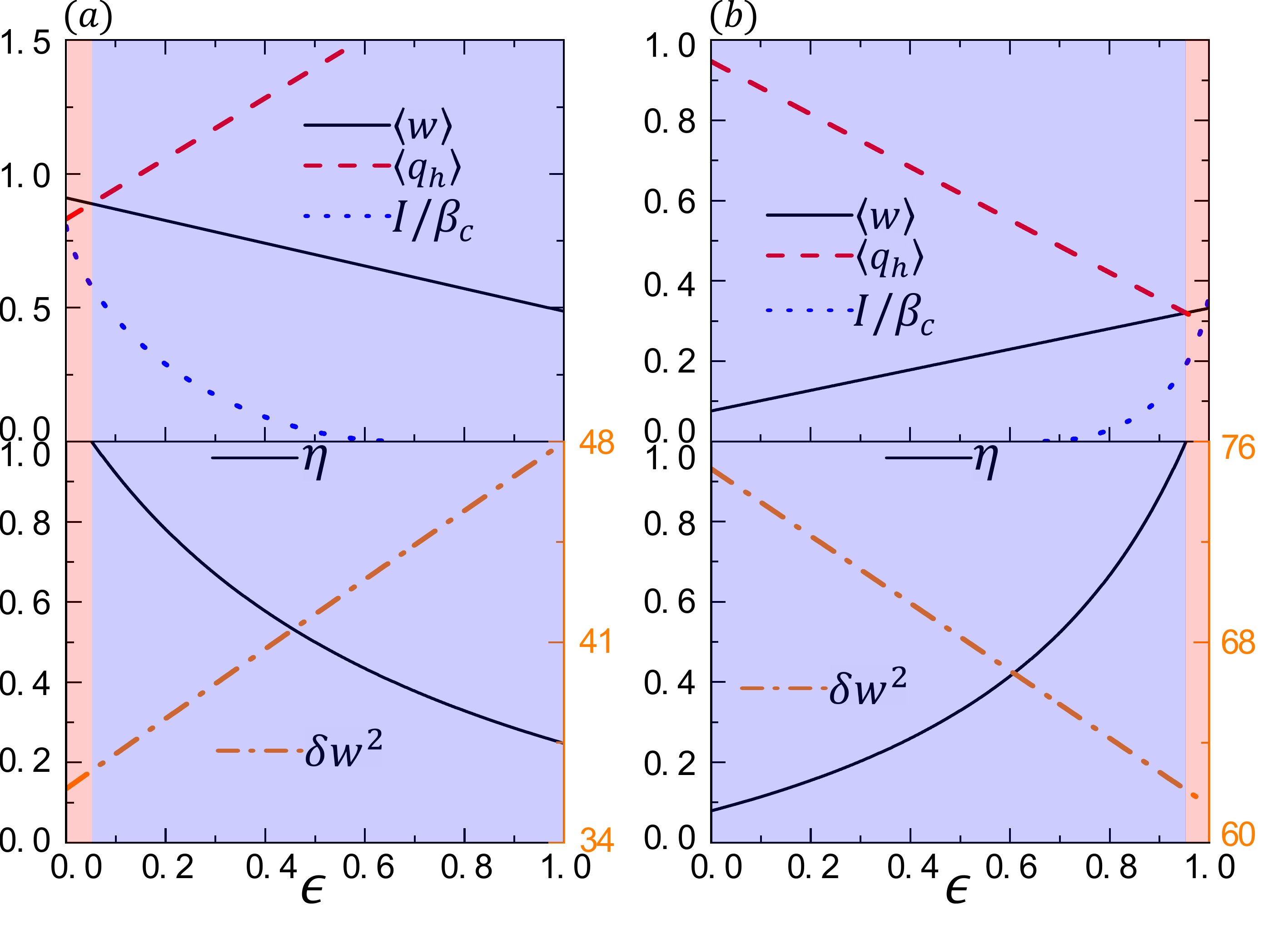}
\end{overpic}
\caption{The  output work $\langle w\rangle$,  the heat $\langle q_{h}\rangle$, the work $I/\beta_c$, the {local} efficiency $\eta$, and the work fluctuations $\delta w^2$ of the dual Otto cycles three-level information engine as  functions of the measurement error $\epsilon$. The right axes of the upper  panels depict the scales for the work fluctuations. {The  blue,  and  red regions represent the regime where $0<\eta<1$, and $\eta=1$ respectively.} In (a) the parameters are  $E_g = 28, E_{e_1} = 11$, and $E_{e_2} = 9$.   In (b) the parameters are   $E_g = 13, E_{e_1} = 26$, and $E_{e_2}= 23$. The other parameters are  $E_c=5$, $\beta_c=0.45$, and $\beta_h=0.1$.}\label{thre}
\end{figure*}

Based on these calculations, we have constructed Fig. \ref{thre}, which  is similar to Fig. \ref{effpe}, therefor indicating  our results can be apploable to the multi-level system. Moreover, given that the information fluctuation theorem and the information dissipation fluctuation theorem are universal principles, the {local} efficiency boundaries are also universal. Thus, we do not provide the boundaries of {local} efficiency in this section.

\section{The dual Otto cycles muti-level TRIE}\label{HO}
{In this section, we  design a dual Otto cycles TRIE based on the muti-level system. Initially, a system with $N+1$ discrete energy levels ($x_i = 0, 1, \dots, N$) is in thermal equilibrium with a cold reservoir at inverse temperature $\beta_c$.  The thermal occupation probability is given by
$p(x_i) = \exp(-\beta_c E_c x_i)/Z_c$ with partition function $Z_c = \sum_{x_i=0}^{N} \exp(-\beta_c E_c x_i),$
where $E_c$ denotes the initial energy gap between adjacent levels. Correspondingly, the average population of the system can be expressed $\langle n_c\rangle=\sum_{x_i=0}^Np(x_i)x_i$.}

{A demon then performs a measurement on the system state, and obtains outcome $y \in \{0, 1, \dots, N\}$ with measurement accuracy 
$P(y = x_i | x_i) = 1 - \epsilon.$
The error distribution is state-dependent:
\begin{align}
    & p(y = j | x_i = 0) = \epsilon/N \quad (j \neq 0), p(y = 0 | x_i \neq 0) = \epsilon/2 \\
   &p(y \neq x_i | x_i \neq 0) = \epsilon/[2(N-1)] \quad \text{for} \quad y \neq 0, x_i
\end{align}
}

{Based on the measurement outcome $y$, the demon controls the subsequent thermodynamic cycle following a protocol analogous to the dual Otto cycles TRIE. When the measurement outcome indicates that the system is in the ground state $y=0$, the system first undergoes adiabatic compression, during which the energy gap between adjacent levels increases from the initial value  $E_c$ to $E_g$. The system then couples to a hot reservoir at an inverse temperature $\beta_h$  during an isochoric heating process until thermal equilibrium is attained, characterized by the probability distribution $p(x_C|y=0)=\exp(-\beta_h E_g x_C)/Z_g$ where $Z_g=\sum_{x_C=0}^N\exp(-\beta_h E_gx_C)$ denotes the partition function. At this stage, the system's average population is given by $\langle n_g\rangle=\sum_{x_C=0}^N p(x_C|y=0)x_C$. Following thermalization, the system undergoes adiabatic expansion, during which the energy gap returns to its initial value $E_c$. Finally, the system couples to the cold reservoir with the fixed energy gap $E_c$ to restore the initial thermal distribution.}

{When the measurement outcome indicates an excited state $y> 0$, the system similarly undergoes adiabatic compression, during which the energy gap increases from $E_c$ to $E_e$.  The system then couples to the hot reservoir until thermal equilibrium is reached, which is now described by the distribution $p(x_C|y)=\exp(-\beta_h E_e x_C)/Z_e$ with  the partition function $Z_e=\sum_{x_C=0}^N\exp(-\beta_h E_ex_C)$. In this case, the average population of the system  is $\langle n_e\rangle=\sum_{x_C=0}^Np(x_C|y)x_C$. The cycle is completed with an adiabatic expansion that restores the energy gap to $E_c$,  followed by an isochoric cooling stroke achieved by recoupling to the cold reservoir, returning the system to the initial thermal distribution.}

{After completing the one cycle, the work $\langle w_{cor}\rangle$ and $\langle w_{err}\rangle$,  the heat $\langle q_{cor}\rangle$ and  $\langle q_{err}\rangle$, and the average of the squared  work $\langle w_{cor}^2\rangle$ and $\langle w_{err}^2\rangle$,  are calculated as
\begin{eqnarray}\label{hjk}
     \langle w_{cor}\rangle&=&\frac{1}{1-\epsilon}\sum_{X_F,y}p(X_F|x_i,y)p(y|x_i)p(x_i)w(X_F,y)\delta_{x_i,y}\nonumber\\
     &=&\frac{1}{1-\epsilon}[\sum_{x_C}^Np(x_C|y)p(y=0|x_i=0)p(x_i=0)(E_g-E_c)x_C\nonumber\\
     &+&\sum_{x_i=1,x_C}^Np(x_C|y)p(y=x_i|x_i)p(x_i)(E_e-E_c)(x_C-x_i)]\nonumber\\
   &=&(E_g-E_c)p(x_i=0)\langle n_g\rangle+(E_e-E_c)\sum_{x_i=1}^Np(x_i)(\langle n_e\rangle-x_i)\nonumber\\
     &=&(E_g-E_c)p(x_i=0)\langle n_g\rangle+(E_e-E_c)[\langle n_e\rangle-\langle n_c\rangle-\langle n_e\rangle p (x_i=0)],\\
  \langle w_{err}\rangle&=&\frac{1}{\epsilon}\sum_{X_F,y}p(X_F|x_i,y)p(y|x_i)p(x_i)w(X_F,y)(1-\delta_{x_i,y})\nonumber\\
   &=&\frac{1}{\epsilon}[\sum_{x_i=1,x_C}^Np(x_C|y=0)p(y=0|x_i)p(x_i)(E_g-E_c)(x_C-x_i)\nonumber\\
    &+&\sum_{y=1,x_C}^Np(x_C|y)p(y|x_i=0)p(x_i=0)(E_e-E_c)x_C\nonumber\\
&+&\sum_{x_i=1,x_C,y\neq0}^Np(x_C|y)p(y\neq x_i|x_i)p(x_i)(E_e-E_c)(x_C-x_i)]\nonumber\\
     &=&\frac{(E_g-E_c)\sum_{x_i=1}^Np(x_i)(\langle n_g\rangle-x_i)}{2}+(E_e-E_c)[\frac{\sum_{y=1}^Np(x_i=0)\langle n_e\rangle}{N}\nonumber\\
     &+&\frac{\sum_{x_i=1}^Np(x_i)\sum_{y\neq0,x_i}^N(\langle n_e\rangle-x_i)}{2(N-1)}]\nonumber\\
     &=&\frac{(E_g-E_c)[\langle n_g\rangle-\langle n_c\rangle-p(x_i=0)\langle n_g\rangle]}{2}\nonumber\\
     &+&(E_e-E_c)\{p(x_i=0)\langle n_e\rangle+\frac{\langle n_e\rangle-\langle n_c\rangle-p(x_i=0)\langle n_e\rangle}{2}\},
     \end{eqnarray}}
     \newpage
{ \begin{eqnarray}
     \langle q_{cor}\rangle&=&\frac{1}{1-\epsilon}\sum_{X_F,y}p(X_F|x_i,y)p(y|x_i)p(x_i)q_h(X_F,y)\delta_{x_i,y}\nonumber\\
    &=&\frac{1}{1-\epsilon}[\sum_{x_C}^Np(x_C|y)p(y=0|x_i=0)p(x_i=0)E_gx_C\nonumber\\
    &+&\sum_{x_i=1,x_C}^Np(x_C|y)p(y=x_i|x_i)p(x_i)E_e(x_C-x_i)]\nonumber\\
    &=&E_gp(x_i=0)\langle n_g\rangle+E_e\sum_{x_i=1}^Np(x_i)(\langle n_e\rangle-x_i)\nonumber\\
     &=&E_gp(x_i=0)\langle n_g\rangle+E_e[\langle n_e\rangle-\langle n_c\rangle-\langle n_e\rangle p (x_i=0)],\\   
      \langle q_{err}\rangle&=&\frac{1}{\epsilon}\sum_{X_F,y}p(X_F|x_i,y)p(y|x_i)p(x_i)q_h(X_F,y)(1-\delta_{x_i,y})\nonumber\\
   &=&\frac{1}{\epsilon}[\sum_{x_i=1,x_C}^Np(x_C|y=0)p(y=0|x_i)p(x_i)E_g(x_C-x_i)\nonumber\\
    &+&\sum_{y=1,x_C}^Np(x_C|y)p(y|x_i=0)p(x_i=0)E_ex_C\nonumber\\
&+&\sum_{x_i=1,x_C,y\neq0}^Np(x_C|y)p(y\neq x_i|x_i)p(x_i)E_e(x_C-x_i)]\nonumber\\
     &=&\frac{E_g\sum_{x_i=1}^Np(x_i)(\langle n_g\rangle-x_i)}{2}+\frac{E_e\sum_{y=1}^Np(x_i=0)\langle n_e\rangle}{N}\nonumber\\
    &+&\frac{E_e\sum_{x_i=1}^N\sum_{y\neq0,x_i}^N(\langle n_e\rangle-x_i)}{2(N-1)}\nonumber\\
     &=&\frac{E_g[\langle n_g\rangle-\langle n_c\rangle-p(x_i=0)\langle n_g\rangle]}{2}+E_ep(x_i=0)\langle n_e\rangle\nonumber\\
    &+&\frac{E_e[\langle n_e\rangle-\langle n_c\rangle-p(x_i=0)\langle n_e\rangle]}{2}
    \end{eqnarray}}
    \newpage
   {\begin{eqnarray}
     \langle w_{cor}^2\rangle&=&\frac{1}{1-\epsilon}\sum_{X_F,y}p(X_F|x_i,y)p(y|x_i)p(x_i)w^2(X_F,y)\delta_{x_i,y}\nonumber\\
     &=&\frac{1}{1-\epsilon}[\sum_{x_C}^Np(x_C|y)p(y=0|x_i=0)p(x_i=0)(E_g-E_c)^2x_C^2\nonumber\\
     &+&\sum_{x_i=1,x_C}^Np(x_C|y)p(y=x_i|x_i)p(x_i)(E_e-E_c)^2(x_C-x_i)^2]\nonumber\\
     &=&(E_g-E_c)^2p(x_i=0)\langle n_g^2\rangle+(E_e-E_c)^2\sum_{x_i=1}^Np(x_i)(\langle n_e^2\rangle-2\langle n_e\rangle x_i+x_i^2)\nonumber\\
    &=&(E_g-E_c)^2p(x_i=0)\langle n_g^2\rangle\nonumber\\
    &+&(E_e-E_c)^2[\langle n_e^2\rangle-2\langle n_e\rangle\langle n_c\rangle+\langle n_c^2\rangle-p (x_i=0)\langle n_e^2\rangle],\\
    \langle w_{err}^2\rangle&=&\frac{1}{\epsilon}\sum_{X_F,y}p(X_F|x_i,y)p(y|x_i)p(x_i)w^2(X_F,y)(1-\delta_{x_i,y})\nonumber\\
    &=&\frac{1}{\epsilon}[\sum_{x_i=1,x_C}^Np(x_C|y=0)p(y=0|x_i)p(x_i)(E_g-E_c)^2(x_C-x_i)^2\nonumber\\
   &+&\sum_{y=1,x_C}^Np(x_C|y)p(y|x_i=0)p(x_i=0)(E_e-E_c)^2x_C^2\nonumber\\
&+&\sum_{x_i=1,x_C,y\neq0}^Np(x_C|y)p(y\neq x_i|x_i)p(x_i)(E_e-E_c)^2(x_C-x_i)^2]\nonumber\\
     &=&\frac{(E_g-E_c)^2\sum_{x_i=1}^Np(x_i)(\langle n_g^2\rangle-2 \langle n_g\rangle x_i+x_i^2)}{2}\nonumber\\
     &+&\frac{(E_e-E_c)^2\sum_{y=1}^Np(x_i=0)\langle n_e^2\rangle}{N}\nonumber\\
     &+&\frac{(E_e-E_c)^2\sum_{x_i=1}^Np(x_i)\sum_{y\neq0,x_i}^N(\langle n_e^2\rangle-2\langle n_e\rangle x_i+x_i^2)}{2(N-1)}\nonumber\\
     &=&\frac{(E_g-E_c)^2[\langle n_g^2\rangle-2\langle n_g\rangle\langle n_c\rangle+\langle n_c^2\rangle-p(x_i=0)\langle n_g^2\rangle]}{2}\nonumber\\
    &+&(E_e-E_c)^2p(x_i=0)\langle n_e^2\rangle\nonumber\\
    &+&\frac{(E_e-E_c)^2[\langle n_e^2\rangle-2\langle n_e\rangle\langle n_c\rangle+\langle n_c^2\rangle-p(x_i=0)\langle n_e^2\rangle]}{2},\label{dfg}
\end{eqnarray}
}     
    \begin{figure*}
 \centering
 \begin{overpic}[width=\linewidth]{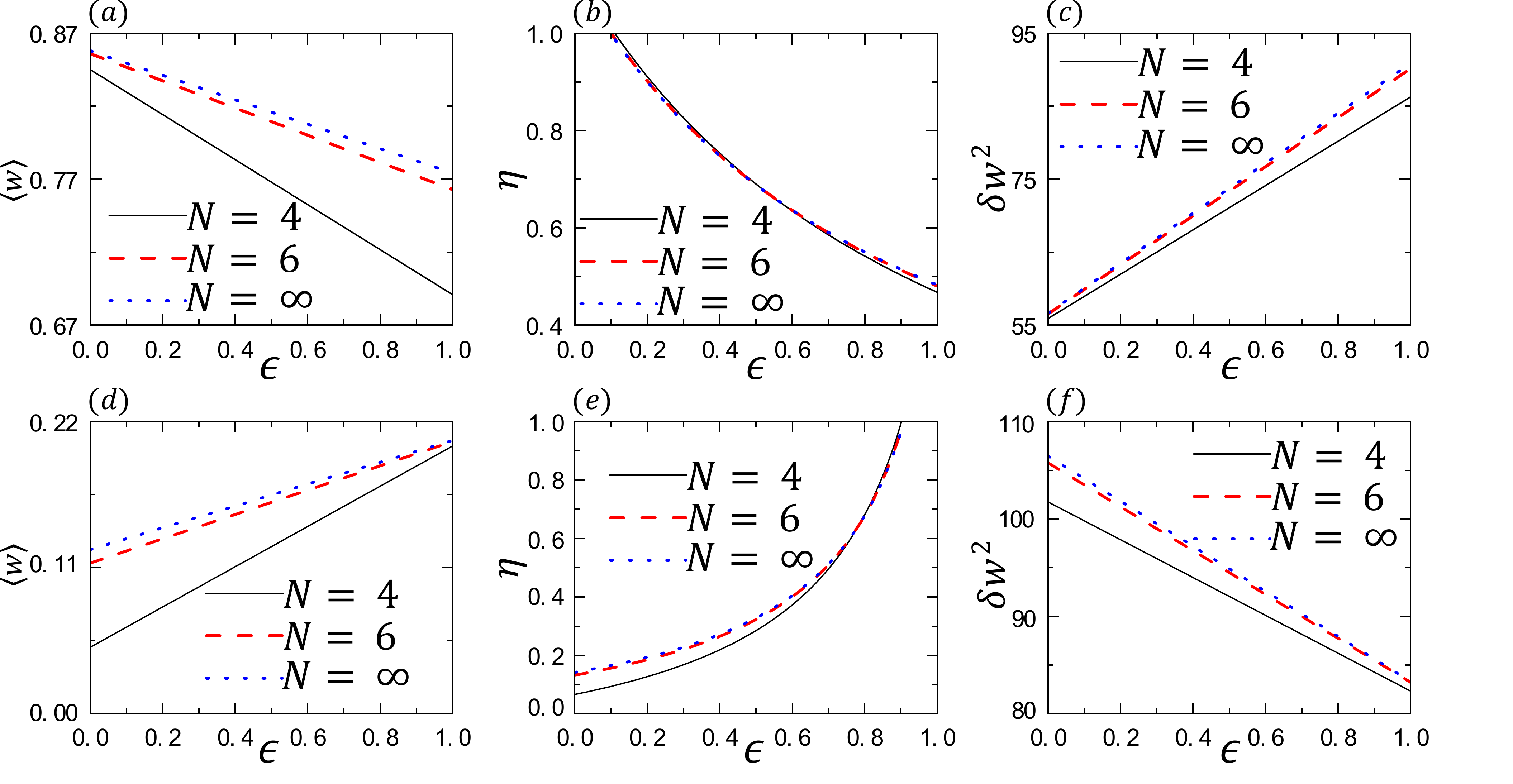}
\end{overpic}
\caption{{The total output work $\langle w\rangle$, the {local} efficiency $\eta$, and the work fluctuations $\delta w^2$   of the dual Otto cycles  TRIE as functions of the measurement error $\epsilon$ with given $N$.  In (a)-(c) the parameters are  $E_g = 22$, and  $E_{e} = 12$.   In (d)-(f) the parameters are   $E_g = 12$, and $E_{e} = 20$. The other parameters are  $E_c=4$, $\beta_c=0.45$, and $\beta_h=0.1$.}}\label{ho}
\end{figure*}
{In the limit $N \to \infty$, this multilevel system reduces to a harmonic oscillator with the ground state energy set to zero. The system's average populations  $\langle n_c\rangle$,  $\langle n_g\rangle$, and $\langle n_e\rangle$  can be expressed as $\langle n_c\rangle=1/[\exp(\beta_c E_c)-1]$, $\langle n_g\rangle=1/[\exp(\beta_h E_g)-1]$, and $\langle n_e\rangle=1/[\exp(\beta_h E_e)-1]$, respectively.  
The average of the square of the population $\langle n_{c,g,e}^2\rangle$  can be determined by the relation  $\langle n_{c,g,e}^2\rangle=\langle n_{c,g,e}\rangle+2\langle n_{c,g,e}\rangle^2$. 
The initial occupancy probability of the ground state is $p_g=1-\exp(-\beta_c E_c)$. 
According to Eqs. (\ref{hjk})-(\ref{dfg}), we have constructed Fig. \ref{ho}, which have the same varying trend as Fig. \ref{effpe}, further indicating that our results are universal. Furthermore, Fig. \ref{ho} also shows that increasing the number of energy levels can increase the output work of the TRIEs.}

\end{appendices}


\newpage


\begin{thebibliography}{99}
\bibitem{Sei12} Bauer, M.,  Abreu, D.,  Seifert, U.:   Efficiency of a Brownian information
 machine. J. Phys. A Math. Theor. \textbf{45}, 162001 (2012).
 \bibitem{Sagawa20}Paneru, G., Dutta, S.,  Sagawa, T.,  Tlusty, T.,  Pak, H. K.:  Efficiency fluctuations and noise induced refrigerator-to-heater transition in information engines, Nat. Commun. \textbf{11}, 1012 (2020).
\bibitem{Sag14}Koski, J. V.,  Maisi, V. F.,  Sagawa, T.,   Pekola, J. P.:  Experimental observation of the role of mutual information in the nonequilibrium dynamics of a Maxwell demon, Phys. Rev. Lett. \textbf{113}, 030601 (2014).
 \bibitem{Yan24} Yan, L. L.,  Bu, J. T.,  Zeng, Q.,  Zhang, K.,  Cui, K. F.,  Zhou, F.,  Su, S., Chen,  L. L., Wang, J.,  Chen, G.,  Feng, M.:  Experimental verification of demon-involved fluctuation theorems,  Phys. Rev.
Lett. \textbf{133}, 090402 (2024).
\bibitem{Szilard1929}Szilard, L.:  $\ddot{u}$ber die entropieverminderung in einem thermodynamischen system bei 
eingriffen intelligenter, Wesen. Z. Physik \textbf{53}, 840-856 (1929).
\bibitem{TK23}Saha, T. K., Ehrich, J., Gavrilov, M.,   Still, S.,  Sivak, D. A.,   Bechhoefer, J.:   Information engine in a nonequilibrium Bath, Phys. Rev. Lett. \textbf{131}, 057101 (2023).
\bibitem{CE18}Elouard, C., Jordan, A. N.:  Efficient quantum measurement
engines, Phys. Rev. Lett. \textbf{120}, 260601 (2018).
\bibitem{Lutz23}Fadler, P.,  Friedenberger, A.,  Lutz, E.:  Efficiency at maximum power of a Carnot quantum information engine, Phys. Rev. Lett. \textbf{130}, 240401 (2023).
\bibitem{ueda18}Masuyama, Y.,  Funo, K.,  Murashita, Y.,  Noguchi, A.,  Kono1, S.,  Tabuchi, Y.,   Yamazaki, R.,  Ueda, M.,  
 Nakamura, Y.:   Information-to-work conversion by Maxwell’s demon in a
 superconducting circuit quantum electrodynamical system,  Nat. Commun. \textbf{9}, 1291 (2018).
 \bibitem{Sagawa08}Sagawa, T., Ueda, M.:  Second law of thermodynamics with discrete quantum feedback control, Phys. Rev. Lett. \textbf{100}, 080403 (2008).
   \bibitem{LC24}Cocconi, L.,  Chen, L.:  Efficiency of an autonomous, dynamic information engine operating on a single active particle, Phys. Rev. E \textbf{110}, 014602(2024).
\bibitem{PM22}Malgaretti, P.,  Stark, H.:   Szilard engines and information-based work extraction for active systems, Phys. Rev. Lett. \textbf{129}, 228005 (2022).
\bibitem{TK22}Saha, T. K., Lucero, J. N. E., Ehrich, J.,  Sivak, D. A., Bechhoefer, J.:  Bayesian information engine that optimally exploits noisy measurements, Phys. Rev. Lett. \textbf{129}, 130601 (2022).
\bibitem{AT22} Taghvaei, A.,  Miangolarra, O. M.,  Fu, R.,  Chen, Y.,  Georgiou, T. T.:
  On the relation between information and power in  stochastic thermodynamic engines, IEEE Control Syst. Lett. \textbf{6}, 434 (2022).

 \bibitem{JD24}du Buisson, J.,  Sivak, D. A.,   Bechhoefer, J.:  Performance limits of information engines, Adv. Phys. X 9, 2352112 (2024).


\bibitem{Landauer61} Landauer, R.:  Irreversibility and heat generation in the computing Process, IBM J. Res. Dev. \textbf{5}, 183 (1961).
\bibitem{KM09}Maruyama, K.,  Nori, F.,  Vedral, V.:  Colloquium: The
 Physics of Maxwell’s Demon and Information, Rev. Mod. Phys. \textbf{81}, 1 (2009).
 \bibitem{Zeng21}Zeng, Q.,  Wang, J.:  New fluctuation theorems on Maxwell’s demon, Sci. Adv. \textbf{7}, 23 (2021). 
\bibitem{EM21}Mizraji, E.:   The biological Maxwell's demons: exploring ideas about the information processing in biological systems. Theory Biosci. \textbf{140}, 307 (2021).
\bibitem{PM12}Mehta, P.,   Schwab, D. J.:  Energetic costs of cellular computation, Proc. Natl. Acad. Sci. U.S.A.
 \textbf{109}, 17978 (2012).
 \bibitem{Lan16} Lan, G., Tu, Y.:  Information processing in bacteria:
 memory, computation, and statistical physics: A key issues
 review, Rep. Prog. Phys. \textbf{79}, 052601 (2016).
 \bibitem{GT16}Tka\v{c}ik, G.,   Bialek, W.:  Information processing in living systems, Annu. Rev. Condens. Matter Phys. \textbf{7}, 89 (2016).
 \bibitem{Wen18}Wen, Y.:  Maxwell’s demon at work: mitochondria, the organelles that convert in
formation into energy? Chronic Dis. Transl. Med. \textbf{4}, 135 (2018).
 \bibitem{Tu08}Tu, Y. H.:  The nonequilibrium mechanism for ultrasensitivity in  a biological switch: Sensing by Maxwell’s demons,  Proc. Natl. Acad. Sci. USA \textbf{105} 11737 (2008).
\bibitem{JS14}Selimkhanov, J.,  Taylor, B.,  Yao, J.,  Pilko, A.,  Albeck, J., Hoffmann, A., Tsimring, L.,  and Wollman, R.:  Accurate information transmission through dynamic biochemical signaling networks, Science \textbf{346}, 1370 (2014).
 \bibitem{HD24}Jiang, H. D.,  Wang, S. S.:  Physical extraction of antigen and information, Proc. Natl. Acad. Sci. U. S. A.,121(39), e2320537121 (2024). 
 \bibitem{Ito15}Ito, S., Sagawa, T.:   Maxwell’s demon in biochemical signal transduction with feedback loop. Nat. Commun. \textbf{6}, 7498 (2015).
\bibitem{tu12}Lan, G.,  Sartori, P.,  Neumann, S.,  Sourjik, V.,  Tu, Y.:  The energy–speed–accuracy trade-off in sensory adaptation, Nat. Phys. \textbf{8}, 422 (2012).
 \bibitem{IL10}Lestas, I.,  Vinnicombe, G.,  Paulsson, J.:  Fundamental limits on the suppression of
 molecular fluctuations, Nature \textbf{467}, 174 (2010).
 \bibitem{Sc24} Schirber, M.:  Information Flow in Molecular Machines, Physics \textbf{17}, 162 (2024).
\bibitem{SF23} Flatt, S.,  Busiello, D. M., Zamuner, S.,  De Los Rios, P.:  ABC transporters are billion-year-old Maxwell
 Demons, Commun Phys \textbf{6}, 205 (2023).
 \bibitem{VS07}Serreli, V.,  Lee, C. F., Kay, E. R.,  and Leigh, D. A.:  A molecular information ratchet, Nature \textbf{445}, 523 (2007). 
 \bibitem{AP2008}
Alvarez-P\'erez, M., ~Goldup, S.~M., ~Leigh, D.~A.,  ~Slawin, A.~M.~Z.:
A Chemically-Driven Molecular Information Ratchet,
J.~Am.~Chem.~Soc.~\textbf{130}, 1836 (2008).
\bibitem{Sag10}Toyabe, S.,  Sagawa, T., Ueda, M.,  Muneyuki, E.,  Sano, M.:  Experimental demonstration of information-to-energy conversion and validation of the generalized Jarzynski equality,
Nat. Phys. \textbf{6}, 988 (2010).
\bibitem{GP18} Paneru, G., Lee, D. Y., Tlusty, T.,  Pak, H. K.:  Lossless Brownian information engine, Phys. Rev. Lett. \textbf{120}, 020601 (2018).
\bibitem{SS20}Seah, S.,  Nimmrichter, S.,  Scarani, V.:  Maxwell’s Lesser Demon: A Quantum Engine Driven by Pointer Measurements, Phys. Rev. Lett.  \textbf{124}, 100603 (2020).
\bibitem{KC17}Chida, K.,  Desai, S.,  Nishiguchi, K.,  Fujiwara, A.:  Power generator driven by Maxwell’s demon, Nat Commun \textbf{8}, 15301 (2017).
\bibitem{JV15}Koski, J. V.,  Kutvonen, A.,  Khaymovich, I. M.,  Ala-Nissila, T.,  Pekola, J. P.:  On-Chip Maxwell’s Demon as an Information-Powered Refrigerator, Phys. Rev. Lett. \textbf{115}, 260602 (2015).
\bibitem{JV14} Koski, J. V.,  Maisi, V. F., Pekola, J. P.,  Averin, D. V.:  Experimental realization of a Szilard engine with a single electron, Proc.
Natl. Acad. Sci. USA \textbf{111}, 13786 (2014).
\bibitem{MD16}Vidrighin, M. D.,  Dahlsten, O.,  Barbieri, M.,  Kim, M. S.,  Vedral, V.,  Walmsley, I. A.:  Photonic Maxwell’s Demon, Phys. Rev. Lett. \textbf{116}, 050401 (2016).
\bibitem{NC17} Cottet, N.,  Jezouin, S., Bretheau, L.,  Campagne-Ibarcq, P.,  Ficheux, Q.,
 Anders, J., Auff$\grave{e}$ves, A.,  Azouit, R.,  Rouchon, P.,  Huard, B.:  Observing a quantum Maxwell demon at work, Proc. Natl. Acad. Sci. U.S.A. \textbf{114}, 7561 (2017).
\bibitem{Naghiloo18}Naghiloo, M., Alonso, J. J., Romito, A., Lutz, E., Murch, K. W.: Information Gain and Loss for a Quantum Maxwell’s Demon,  Phys. Rev. Lett. \textbf{121}, 030604 (2018).
\bibitem{JH14} Hoppenau J. Engel, A.:  On the energetics of information exchange, Europhys. Lett. \textbf{105}, 50002 (2014).
\bibitem{JH2214} Um, J.,  Hinrichsen, H.,  Kwon, C.,  Park, H.:  Total cost of operating an information engine, New J. Phys. \textbf{17}, 085001 (2015).
\bibitem{PS14}Pal, P. S.,  Rana, S., Saha,  A.,  Jayannavar, A. M.:  Extracting work from a single heat bath:
A case study of a Brownian particle under an external magnetic field in the presence of information, Phys. Rev. E \textbf{90}, 022143 (2014).
\bibitem{DAS24}Johan du Buisson, D. A. S., Bechhoefer, J.:  Performance limits of information engines, Adv. Phys.: X \textbf{9},
 2352112 (2024).
 \bibitem{KV25}V. K.,  Joseph, T.:  Imprecise Maxwell’s demon with feedback delay: An exactly solvable information engine model,  Phys. Rev. E \textbf{111}, L042104 (2025).
\bibitem{NE25} Navid Elyasi, S.,  Rossi, M. A. C.,  Genoni, M. G. :
Experimental simulation of daemonic work extraction in
open quantum batteries on a digital quantum computer,
Quantum Science and Technology \textbf{10}, 025017 (2025).
\bibitem{JP23}Garrahan, J. P.,  Ritort, F. :  Generalized continuous Maxwell demons. Phys. Rev. E \textbf{107}, 034101 (2023).
\bibitem{MR19} Ribezzi-Crivellari, M.,  Ritort, F. : Large work extraction and the Landauer limit in a continuous Maxwell demon. Nature Physics \textbf{15}, 660 (2019).
\bibitem{Seifert11}  Abreu, D. ,  Seifert, U. : Extracting work from a single heat bath through feedback,  Europhys. Lett. \textbf{94}, 10001 (2011).
\bibitem{KT25} Kamijima, T.,    Takatsu, A., 
  Funo, K.,
  Sagawa, T.: Optimal finite-time Maxwell’s demons in Langevin systems, Phys. Rev. Res. \textbf{7}, 023159 (2025).
\bibitem{JM11} Horowitz. J. M.,   Parrondo, J. M. R. : Designing optimal discrete-feedback thermodynamic engines, New J. Phys. \textbf{13}, 123019 (2011).
\bibitem{Ueda12} Sagawa, T.,  Ueda, M. : Nonequilibrium thermodynamics of feedback control,  Phys. Rev. E \textbf{85}, 021104 (2012).
\bibitem{hol18} Holubec, V.,   Ryabov, A.:  Cycling Tames Power Fluctuations near Optimum Efficiency, Phys. Rev. Lett.  \textbf{121}, 120601 (2018).
\bibitem{sei18}Pietzonka, P.,  Seifert, U.:  Universal Trade-Off between Power, Efficiency, and Constancy in Steady-State Heat Engines, Phys. Rev. Lett. \textbf{120}, 190602 (2018).
\bibitem{lutz21} Bouton, Q.,  Nettersheim, J.,  Burgardt, S.,  Adam1, D.,  Lutz, E.,   Widera, A.:  A quantum heat engine driven by atomic collisions, Nat. Commun. \textbf{12}, 2063 (2021).
 \bibitem{Quan07}Quan, H. T.,  Liu, Y. X., Sun, C. P.,  Nori, F.:   Quantum thermodynamic cycles and quantum heat engines, Phys. Rev. E \textbf{76}, 031105 (2007).
 \bibitem{TD2004}Kieu, T. D.:  The second law, Maxwell's demon, and work derivable from quantum heat engines, Phys. Rev. Lett. \textbf{93}, 140403 (2004).
\bibitem{sagawa12}Sagawa, T., Ueda,  M.: Fluctuation theorem with information exchange: role of correlations in stochastic thermodynamics,  Phys. Rev. Lett. \textbf{109}, 180602 (2012).
\bibitem{TT25}Tohme, T.,  Bedoya, V.,  di Bello, C.,  Bresque, L., Manzano, G.,  and  Roldán, É.: Gambling Carnot Engine, Phys. Rev. Lett.  \textbf{135}, 067101 (2025).
\bibitem{Jar13}Deffner, S.,  Jarzynski, C.:  Information Processing and the Second Law of Thermodynamics:
An Inclusive, Hamiltonian Approach,  Phys. Rev. X, \textbf{3}, 041003 (2013).
 \bibitem{Crook99} Crooks, G. E.: Entropy production fluctuation theorem and the nonequilibrium work  relation for free energy differences. Phys. Rev. E \textbf{60}, 2721 (1999).
\bibitem{Jar00}Jarzynski, C.:  Hamiltonian derivation of a detailed fluctuation theorem. J. Stat. Phys. \textbf{98}, 77 (2000).
 \bibitem{Zhang22}Zhang, K.,  Wang, X.,  Zeng, Q.,  Wang, J.:  Conditional
 entropy production and quantum fluctuation theorem of
 dissipative information: Theory and experiments, PRX
 Quantum \textbf{3}, 030315 (2022).
 \bibitem{Lutz24}Hou, W. E.,  Zhao, X. Y., Rehan, K.,  Li, Y.,  Li, Y.,
 Lutz, E.,  Lin, Y. H.,  Du, J. F.:  Combining energy efficiency and quantum advantage in cyclic machines,  Nat Commun \textbf{16}, 5127 (2025).
\bibitem{OV23}Ivakhnenko, O. V.,  Shevchenko, S. N.,  Nori, F.: 
Nonadiabatic Landau-Zener-St$\ddot{\mathrm{u}}$ckelberg-Majorana transitions, dynamics, and interference, Phys. Rep. \textbf{995}, 1 (2023).
\bibitem{JP29} Peterson, J. P. S.,  Batalhão, T. B.,  Herrera, M.,  Souza, A. M., 
Sarthour, R. S.,   Oliveira, I. S., Serra, R. M.:  Experimental
characterization of a spin quantum heat engine, Phys. Rev. Lett.
\textbf{123}, 240601 (2019).
\bibitem{Yan22} Yan, L. L.,   et al.: Experimental
verification of dissipation-time uncertainty relation, Phys. Rev.
Lett. \textbf{128}, 050603 (2022).
\bibitem{TM60}Cover, T. M.,  Thomas, J. A.:  \textit{Elements of information theory} (Wiley, ed. 2, 2006).
{\bibitem{Zhang2025}  Zhang, J. et al. Single-ion information engine for charging quantum battery. Phys. Rev. Lett. 135, 140403 (2025).}
\end{thebibliography}
\end{document}